\documentclass[12pt]{article}
\usepackage{epsfig}\parskip 5pt plus 1pt
\newcommand{\nn}{\nonumber}

\newcommand{\be}{\begin{equation}}
\newcommand{\ee}{\end{equation}}
\newcommand{\bea}{\begin{eqnarray}}
\newcommand{\eea}{\end{eqnarray}}

\begin{document}
%
\thispagestyle{empty}
\begin{flushright}
{\tt hep-ph/0305185}\\
\today
\end{flushright}
\vspace*{1cm}

\begin{center}
{\Large{\bf The synergy of the golden and silver channels at the
Neutrino Factory } }\\
\vspace{.5cm} 

D.~Autiero$^{\rm a}$, G.~De~Lellis$^{\rm b}$, A.~Donini$^{\rm c}$,
M.~Komatsu$^{\rm d}$, D.~Meloni$^{\rm e}$, P.~Migliozzi$^{\rm f}$,
R.~Petti$^{\rm g}$, L.~Scotto~Lavina$^{\rm b}$ and F.~Terranova$^{\rm
h}$
 
\vspace*{1cm} $^{\rm a}$ IPN Lyon, F-69622, Villeurbanne (Lyon),
France\\ $^{\rm b}$ Dip. Fisica, Universit\`a di Napoli``Federico II''
and I.N.F.N., I-80126, Napoli, Italy \\ $^{\rm c}$ I.F.T. and
Dep. F\'{\i}sica Te\'orica, U.A.M., E-28049, Madrid, Spain \\ $^{\rm
d}$ Nagoya University, Nagoya 464-8602, Japan \\ $^{\rm e}$
Dep. F\'{\i}sica Te\'orica y del Cosmos, Univ. Granada, E-18002,
Granada, Spain \\ $^{\rm f}$ I.N.F.N., Sezione di Napoli, I-80126,
Napoli, Italy \\ $^{\rm g}$ C.E.R.N., CH-1211, Geneva, Switzerland \\
$^{\rm h}$ I.N.F.N., Laboratori Nazionali Frascati, I-00044, Frascati
(Rome), Italy \\
\end{center}

\vspace{.3cm}

\begin{abstract}
\noindent
We deepen the study of the so-called ``silver channel'' $\nu_e \to
\nu_\tau$~\cite{Donini:2002rm} and of its relevance to solve some of
the ambiguities that can arise in the simultaneous measurement of
$(\theta_{13},\delta)$ at the Neutrino Factory by presenting in full
detail the characteristics of the considered OPERA-like detector and
the experimental treatment of the different backgrounds and signals.
Furthermore, we perform a detailed study of the systematic errors
associated both with the OPERA-like and the magnetized-iron detectors
and their effects on the sensitivity. Finally, we also apply a refined
statistical analysis of the simulated events based on the frequentist
approach.
\end{abstract}

\newpage

\section{Introduction}

The hypothesis of neutrino oscillations~\cite{Pontecorvo:1957yb} is at
present strongly supported by atmospheric, solar,
accelerator~\cite{evidence_osc} and reactor~\cite{KAMLAND} neutrino
data. If we do not consider the claimed evidence for oscillations at
the LSND experiment~\cite{Athanassopoulos:1998pv}, that must be
confirmed or excluded by the ongoing MiniBooNE
experiment~\cite{zimmermann}, oscillations in the leptonic sector can
be easily accommodated in the three family
Pontecorvo-Maki-Nakagawa-Sakata (PMNS) mixing matrix $U_{PMNS}$:
\begin{eqnarray}
\label{pmns}
 U_{PMNS} =
\left ( 
\begin{array}{ccc}
1 &        0 &       0 \\
0 &   c_{23} &  s_{23} \\
0 & - s_{23} &  c_{23} 
\end{array}
\right ) \;
\left (
\begin{array}{ccc}
c_{13} & 0 &  s_{13} e^{i \delta} \\
0 & 1 & 0 \\
-  s_{13} e^{-i \delta} & 0 &  c_{13}
\end{array}
\right ) \;
\left (
\begin{array}{ccc}
 c_{12} &  s_{12} & 0 \\
- s_{12} &  c_{12} & 0 \\
0 & 0 & 1
\end{array}
\right ) \, , \\
\nn
\end{eqnarray}
with the short-form notation $ s_{ij} \equiv \sin \theta_{ij}, c_{ij}
\equiv \cos \theta_{ij}$.  Further Majorana phases have not been
introduced, since oscillation experiments are only sensitive to the
two neutrino mass differences $\Delta m^2_{12}, \Delta m^2_{23}$ and
to the four parameters in the mixing matrix of eq.~(\ref{pmns}): three
angles and the Dirac CP-violating phase, $\delta$.

In particular, data from atmospheric neutrinos and from K2K are
interpreted as oscillations of $\nu_\mu$ into $\nu_\tau$ with a mass
gap that we denote by $\Delta m^2_{23}$.  The corresponding mixing
angle is close to maximal, $\sin^2 2 \theta_{23} > 0.9$, and $|\Delta
m^2_{23}|$ is in the range $1.9$ to $3.5 \times 10^{-3}$
eV$^2$~\cite{yanagisawa}.  On the other hand, the longstanding solar
neutrino problem has finally been solved by the combination of the SNO
data~\cite{Ahmad:2002jz} and recent KamLand results~\cite{KAMLAND},
isolating the LMA-MSW region~\cite{MSW} as the only viable solution of the
solar neutrino deficit with $\nu_e$ oscillations into active
($\nu_\mu, \nu_\tau$) neutrino states. Two allowed regions are
identified inside the LMA-MSW solution~\cite{Fogli:2002au},
corresponding to $\Delta m^2_\odot =\Delta m^2_{12} \simeq 7 \times 10^{-5}$ eV$^2$
(LMA-I) and $\Delta m^2_{12} \simeq 1.4 \times 10^{-4}$ eV$^2$
(LMA-II).  For both solutions the corresponding mixing angle
($\theta_{12}$) is large (albeit maximal mixing is excluded at $3\sigma$).
Finally, a comprehensive three-family analysis (including the negative
CHOOZ results~\cite{chooz}) put a bound on $\theta_{13}$,
$\sin^2\theta_{13} \leq 0.02$.

The planned long baseline experiments~\cite{ICARUS}-\cite{JHF} will
improve the measurement of $\Delta m^2_{atm}$ ($\Delta m^2_{atm}
\simeq \Delta m^2_{23} \simeq\Delta m^2_{13}$) and of $\theta_{23}$
and measure or increase the bound on
$\theta_{13}$~\cite{komatsu,minosIN} (see
also~\cite{Migliozzi:2003pw}). This new generation of experiments,
however, is only the first step toward the ambitious goal of precision
measurements of the whole three-neutrino mixing parameter space,
including the leptonic CP-violating phase $\delta$.  This long-lasting
experimental program consists of the development of some ``superbeam''
facilities (whose combination can strongly improve our knowledge on
$\theta_{13}$, see~\cite{Huber:2002mx}) and, eventually, of a
``Neutrino Factory'' (high-energy muons decaying in the straight
section of a storage ring, thus producing a very pure and intense
two-flavor neutrino beam~\cite{Geer:1998iz,DeRujula:1998hd}). One of
the main goals of the Neutrino Factory program (see for example
\cite{Blondel:2000gj,Apollonio:2002en} and refs. therein) would be the
discovery of leptonic CP violation and, possibly, its
study~\cite{Dick:1999ed}-\cite{Cervera:2000kp}.

The most sensitive method to study this topic is the measure of the
transition probability $\nu_e(\bar \nu_e) \rightarrow \nu_\mu(\bar
\nu_\mu)$. This is what is called the ``{\it golden measurement at the
Neutrino Factory}''~\cite{Cervera:2000kp}. 
At the Neutrino Factory an energetic electron neutrino beam is produced 
with no contamination from muon neutrinos with the same helicity (only muon
neutrinos of opposite helicity are present in the beam, contrary to the
case of conventional beams from pion decay). Therefore, the transition
of interest can be easily measured by searching for wrong-sign muons,
i.e. muons with charge opposite to that of the parent muons in the
storage ring, provided the considered detector has a good muon charge 
identification capability. However,
the determination of ($\theta_{13},\delta$) at the Neutrino Factory is
not at all free of ambiguities, much as it was the case for the
different solutions to the solar neutrino problem before KamLand.
In~\cite{Burguet-Castell:2001ez} it was shown that, for a given
physical input parameter pair ($ \theta_{13}, \delta$),
measuring the oscillation probability for $\nu_e \to \nu_\mu$ and
$\bar \nu_e \to \bar \nu_\mu$ will generally result in two allowed
regions of the parameter space. The first one contains the physical
input parameter pair and the second, the ``intrinsic ambiguity'', is
located elsewhere. Worse than that, new degeneracies have later been
noticed~\cite{Minakata:2001qm,Barger:2001yr}, resulting from our
ignorance of the sign of the $\Delta m^2_{atm}$ squared mass
difference and from the approximate $[\theta_{23}, \pi/2 -
\theta_{23}]$ symmetry for the atmospheric angle. In general, for each
physical input pair the measure of $P(\nu_e \to \nu_\mu)$ and $P(\bar
\nu_e \to \bar \nu_\mu)$ will result in eight allowed regions of the
parameter space (if the sign of $\Delta m^2_{atm}$ and the
$\theta_{23}$- octant will still be unknown by the time the Neutrino
Factory will be operational)~\cite{Barger:2001yr}.

Different proposals have been suggested to deal with the three
ambiguities. The sign of $\Delta m^2_{atm}$ could for example be
determined by combining two of the planned superbeam facilities, one
of them with sufficiently long baseline while the second one with good
$\theta_{13}$ sensitivity~\cite{Minakata:2003ca}. The
$\theta_{23}$-octant could be determined by combining one of these
superbeam facilities (e.g. JHF-I) with a reactor-driven
detector~\cite{Minakata:2002jv} (although in this case systematics
could represent a serious issue). Finally, the intrinsic ambiguity can
be solved by fitting at two different baselines at the same
time~\cite{Burguet-Castell:2001ez}; or by increasing the energy
resolution of the
detector~\cite{Freund:2001ui}-\cite{Kajita:2001sb}.
In~\cite{Burguet-Castell:2002qx} a detailed study of the combination
of a superbeam and of a (single-baseline) Neutrino Factory was
presented, showing that this combination is extremely helpful in
solving ambiguities whenever a 1 Mton superbeam-driven water Cerenkov
detector is added to the Neutrino Factory-driven 40 Kton magnetized
iron detector (MID) that was considered in previous
studies~\cite{Cervera:2000vy}.

In all of these proposals only the ``golden channel'' ($\nu_e \to
\nu_\mu$, or $\nu_\mu \to \nu_e$ in the case of superbeams), with
different detectors and neutrino sources, was
considered. However, in~\cite{Donini:2002rm} it was noticed that muons
proceeding from $\tau$ decay when $\tau$'s are produced via a $\nu_e
\to \nu_\tau$ transition show a different $(\theta_{13}, \delta)$
correlation from those coming from $\nu_e \to \nu_\mu$.  By using a
lead-emulsion detector based on the Emulsion Cloud Chamber (ECC)
technique, capable of the $\tau$-decay vertex recognition, it is
therefore possible to use the complementarity of the information from
$\nu_e \to \nu_\tau$ and from $\nu_e \to \nu_\mu$ to solve the
intrinsic ($\theta_{13}, \delta$) ambiguity.  The lesser statistical
significance of the former\footnote{Due to the mass of the ECC
detector, to the $\nu_\tau N$ CC cross-section, to the $BR(\tau \to
\mu)$ factor and to the need for the vertex identification, which
results in a smaller total efficiency.}  with respect to the latter
(the ``golden'' muons), inspired the nickname of ``silver channel''
and consequently of ``silver'' muons. Silver muons, in combination
with golden muons, could also be extremely helpful in dealing with the
$[\theta_{23},\pi/2-\theta_{23}]$ ambiguity, since the leading term in
$P(\nu_e \to \nu_\tau)$ is proportional to $\cos^2 \theta_{23}$,
whereas the analogous term in $P(\nu_e \to \nu_\mu)$ is proportional
to $\sin^2 \theta_{23}$. However, the sensitivity of the silver/golden
channel combination to the $\theta_{23}$-octant strongly depends on
the value of $\theta_{13}$. A detailed study of this combination to
solve the $\theta_{23}$-octant ambiguity is currently underway
(see~\cite{doniniNOON}).  In this paper, as in~\cite{Donini:2002rm},
we restrict ourselves to the ($\theta_{13},\delta$) ambiguity, by
fixing $\theta_{23} = 45^\circ$ and by choosing a given sign for
$\Delta m^2_{atm}$ (in the hypothesis that more information on the
three neutrino spectrum will be available by the time the Neutrino
Factory will be operational). Clearly, solving the three ambiguities
at the same time will need the combination of different measurements.

In~\cite{Donini:2002rm}, a preliminary analysis of the foreseeable
backgrounds in the measurement of $\nu_e \to \nu_\tau$ transitions was
used to substantiate the proposal with a realistic simulation of the
detector based on~\cite{proposal,Guler:2001hz}. The aim of this paper
is to present a dedicated analysis to measure the ``silver'' muon
signal at an OPERA-like lead-emulsion detector. Furthermore, a
detailed study of the systematic errors associated to both ECC and MID
is also presented. In particular, it is shown that, by including a
realistic systematic error in the analysis of the golden channel, its
performance in determining the ($\theta_{13},\delta$) value is
considerably worsened if compared to the ones reported
in~\cite{Cervera:2000vy}. A refined statistical analysis of ``silver''
and ``golden'' muons is also included, in order to optimally deal with
very low statistics signals but with a very high signal-to-noise
ratio. We consider in this analysis two possible combinations of
detectors and baselines:
\begin{itemize}
\item An ECC at $L = 732$ km and a MID at $L = 3000$ km \\
     (same combination considered in~\cite{Donini:2002rm});
\item An ECC at $L = 3000$ km and a MID at $L = 3000$ km .
\end{itemize}

Our conclusion is that the combination of ``golden'' and ``silver''
channel at the Neutrino Factory can indeed solve the intrinsic
ambiguity problem for $\theta_{13} > 1^\circ$, thus confirming the
results of \cite{Donini:2002rm}. For $\theta_{13} = 1^\circ$ we are on
the edge of the experimental sensitivity of the silver channel with
both configurations. However, given the lower expected background, we
find that the second configuration (both detectors at 3000~km) allows
a reduction in the size of the confidence interval for $\theta_{13} >
1^\circ$.

The paper is organized as follows: in Sect.~\ref{sec:channel} we
introduce the ``golden'' and ``silver channels'' at the Neutrino
Factory and we show how a combination of these two signals could solve
the intrinsic ambiguity $(\theta_{13},\delta)$; in
Sect.~\ref{sec:rates} we recall the neutrino fluxes, the $\nu N$
cross-sections and finally report the expected number of events for
each channel; in Sect.~\ref{sec:issues} the characteristics of a
lead-emulsion OPERA-like detector are depicted; in
Sect.~\ref{sec:anasilver} we discuss in full detail the ``silver
muons'' signal and the different sources of background at an ECC
detector; in Sect.~\ref{sec:scanning} we give an estimate of the
required scanning load; in Sect.~\ref{sec:refined} we perform a
refined analysis of low statistics signals with a very good S/N within
the Feldman-Cousins~\cite{Feldman:1997qc} approach; in
Sect.~\ref{sec:concl} we eventually draw our conclusions.

\section{The Golden and the Silver Channel}
\label{sec:channel}

Following eq.~(1) of~\cite{Burguet-Castell:2001ez}, we get for the
transition probability $\nu_e \to \nu_\mu$ ($\bar \nu_e \to \bar
\nu_\mu$) and $\nu_e \to \nu_\tau$ ($\bar \nu_e \to \bar \nu_\tau$) at
second order in perturbation theory in $\theta_{13}$,
$\Delta_\odot/\Delta_{atm}$, $\Delta_\odot/A$ and $\Delta_\odot L$
(see also~\cite{Freund:2001pn}-\cite{Minakata:2002qe}), \be
\label{eq:spagnoli}
P^\pm_{e \mu} (\bar \theta_{13}, \bar \delta) = X_\pm \sin^2 (2 \bar
\theta_{13}) + Y_\pm \cos ( \bar \theta_{13} ) \sin (2 \bar
\theta_{13} ) \cos \left ( \pm \bar \delta - \frac{\Delta_{atm} L }{2}
\right ) + Z \, , \ee and \be
\label{eq:etau}
P^\pm_{e \tau} (\bar \theta_{13}, \bar \delta) = 
X^\tau_\pm \sin^2 (2 \bar \theta_{13}) -
Y_\pm \cos ( \bar \theta_{13} ) \sin (2 \bar \theta_{13} )
      \cos \left ( \pm \bar \delta - \frac{\Delta_{atm} L }{2} \right ) 
+ Z^\tau \, ,
\ee
where $\pm$ refers to neutrinos and anti neutrinos, respectively.
The coefficients of the two equations are:
\be
\label{eq:xcoeff}
\left \{ 
\begin{array}{lll}
X_\pm &=& \sin^2 (\theta_{23} ) 
\left ( \frac{\Delta_{atm} }{ B_\mp } \right )^2
\sin^2 \left ( \frac{ B_\mp L}{ 2 } \right ) \ , \\
\nn \\
X^\tau_\pm &=& \cos^2 (\theta_{23} ) 
\left ( \frac{\Delta_{atm} }{ B_\mp } \right )^2 
\sin^2 \left ( \frac{ B_\mp L}{ 2 } \right ) \ , 
\end{array} 
\right .
\ee
\be
\label{eq:ycoeff}
Y_\pm = \sin ( 2 \theta_{12} ) \sin ( 2 \theta_{23} )
\left ( \frac{\Delta_\odot }{ A } \right )
\left ( \frac{\Delta_{atm} }{ B_\mp } \right )
\sin \left ( \frac{A L }{ 2 } \right )
\sin \left ( \frac{ B_\mp L }{ 2 } \right ) \ , 
\ee
\be
\label{eq:zcoeff}
\left \{ 
\begin{array}{lll}
Z &=& \cos^2 (\theta_{23} ) \sin^2 (2 \theta_{12})
\left ( \frac{\Delta_\odot }{ A } \right )^2
\sin^2 \left ( \frac{A L }{ 2 } \right ) \ , \\
\nn \\
Z^\tau &=& \sin^2 (\theta_{23} ) \sin^2 (2 \theta_{12}) 
\left ( \frac{\Delta_\odot }{ A } \right )^2 
\sin^2 \left ( \frac{A L }{ 2 } \right ) \ ,
\end{array}
\right .  
\ee
with $A = \sqrt{2} G_F n_e$ (expressed in
eV$^2$/GeV) and $B_\mp = | A \mp \Delta_{atm}|$ (with $\mp$ referring
to neutrinos and anti-neutrinos, respectively). Finally, $\Delta_{atm}
= \Delta m^2_{atm} / 2 E_\nu$ and $\Delta_\odot = \Delta m^2_\odot / 2
E_\nu$.

The parameters $\bar \theta_{13}$ and $\bar \delta$ are the physical
parameters that must be reconstructed by fitting the experimental data
with the theoretical formula for oscillations in matter.  In what
follows, the other parameters have been considered as fixed
quantities, supposed to be known with good precision by the time when
the Neutrino Factory will be operational. In particular,
\begin{enumerate} 
\item in the solar sector we fixed $\theta_{12} = 33^\circ$ and
$\Delta m^2_\odot = 1.0 \times 10^{-4} $ eV$^2$ (same as
in~\cite{Donini:2002rm}).  Although these values do not correspond to
the present best-fit value from the combined analysis of the solar and
KamLand data~\cite{Fogli:2002au}, they have been chosen to make direct
comparison with our previous results.  Furthermore, notice that the
squared mass difference lies in between the two allowed solutions
LMA-I with $\Delta m^2_\odot = 7 \times 10^{-5}$ eV$^2$ and LMA-II
with $\Delta m^2_\odot = 1.4 \times 10^{-4}$ eV$^2$~\cite{KAMLAND}.
\item in the atmospheric sector, $\theta_{23} = 45^\circ$ and $\Delta
  m^2_{atm} = 2.9 \times 10^{-3} $ eV$^2$~\cite{yanagisawa}, with
  $\Delta m^2_{atm}$ positive.  Notice that for $\theta_{23} =
  45^\circ$ the $\theta_{23}$-octant ambiguity~\cite{Barger:2001yr} is
  absent.
\end{enumerate}

Finally, we also considered a fixed value for the matter parameter, $A
= 1.1 \times 10^{-4} $ eV$^2$/GeV (this value, obtained by using the
average matter density alongside the path for the chosen distance
computed with the Preliminary Earth Model~\cite{earthmodel}, is a good
approximation for the case under consideration of $L < 4000$ km).  For
simplicity, we have not included errors on these parameters\footnote{
It has been shown in~\cite{Burguet-Castell:2001ez} that the inclusion
of the foreseeable uncertainties on these parameters does not modify
the results on the $\theta_{13}$ and $\delta$ measurements in a
significant way.}.

Eqs.~(\ref{eq:spagnoli}) and~(\ref{eq:etau}) lead to two
equiprobability curves in the ($\theta_{13}, \delta$) plane for
neutrinos and anti-neutrinos of a given energy: \bea
\label{eq:equi0} 
P^\pm_{e \mu} (\bar \theta_{13}, \bar \delta) = P^\pm_{e \mu}
(\theta_{13}, \delta) \, ; \\ \nn \\ P^\pm_{e \tau} (\bar \theta_{13},
\bar \delta) = P^\pm_{e \tau} (\theta_{13}, \delta) \, .  \eea Notice
that $X^\tau_\pm$ and $Z^\tau$ differ from the corresponding
coefficients for the $\nu_e \to \nu_\mu$ transition for the $\cos
\theta_{23} \leftrightarrow \sin \theta_{23}$ exchange, only (and thus
for $\theta_{23} = 45^\circ$ we have $X = X^\tau, Z = Z^\tau$).  The
$Y_\pm$ term is identical for the two channels, but it appears with an
opposite sign. This sign difference in the $Y$-term is crucial, as it
determines a different shape in the $(\theta_{13}, \delta)$ plane for
the two sets of equiprobability curves.

\begin{figure}[h!]
\begin{center}
\begin{tabular}{cc}
\hspace{-1cm} \epsfxsize7cm\epsffile{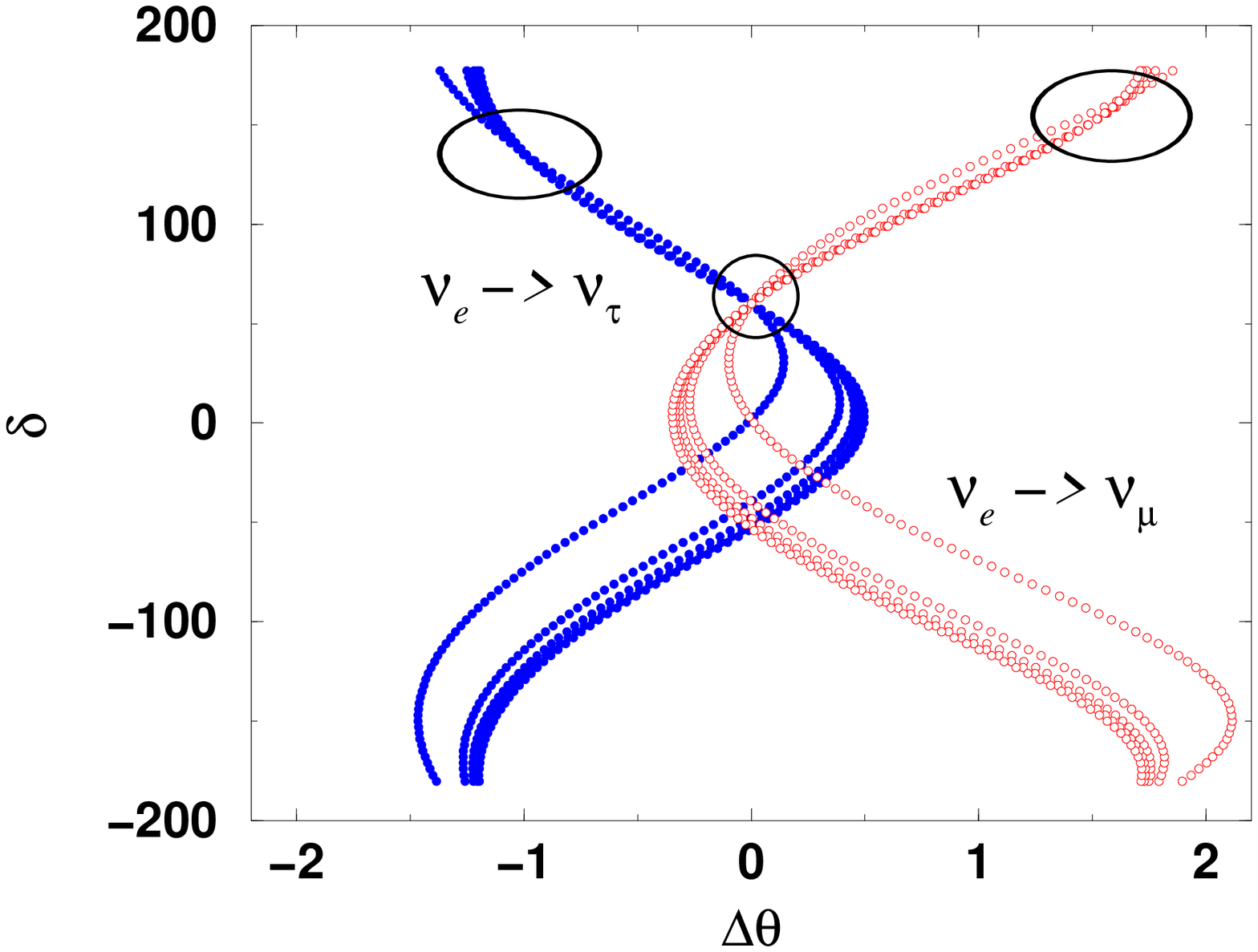} &
              \epsfxsize7cm\epsffile{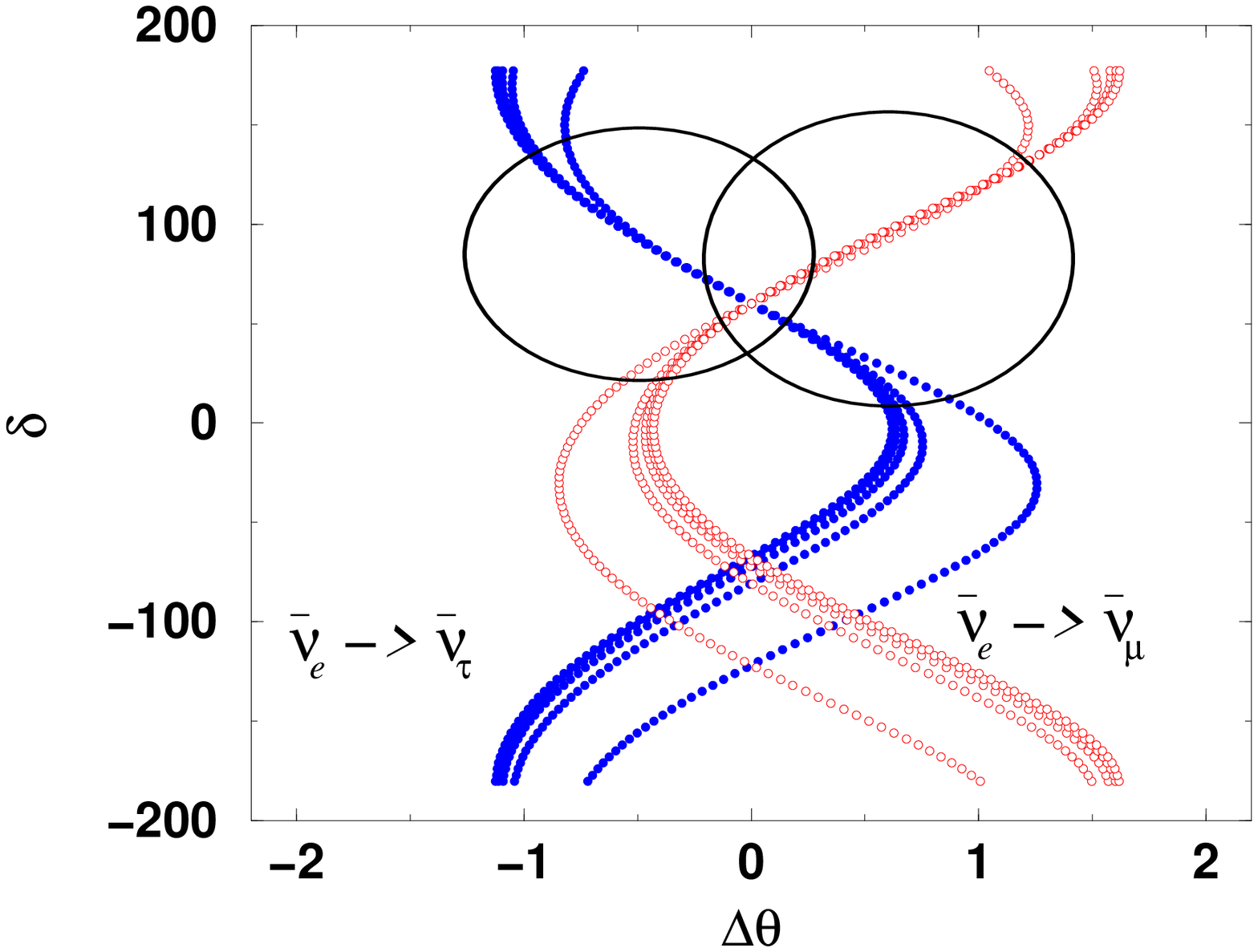}
\end{tabular}
\caption{\it Equiprobability curves in the ($\Delta \theta, \delta$)
plane, for $\bar \theta_{13} = 5^\circ, \bar \delta = 60^\circ$,
$E_\nu \in [5, 50] $ GeV and $L = 732$ km for the $\nu_e \to \nu_\mu$
and $\nu_e \to \nu_\tau$ oscillation (neutrinos on the left,
anti-neutrinos on the right). $\Delta \theta$ is defined as the
difference between the reconstructed parameter $ \theta_{13}$ and the
input parameter $\bar{\theta}_{13}$, $\Delta \theta = 
\theta_{13}- \bar{\theta}_{13}$.}
\label{fig:equiprobetau}
\end{center}
\end{figure}

In Fig.~\ref{fig:equiprobetau}, we superimposed the equiprobability
curves for the $\nu_e \to \nu_\tau$ and $\nu_e \to \nu_\mu$
oscillations at a fixed distance, $L = 732$ km, with input parameters
$\bar \theta_{13} = 5^\circ$ and $\bar \delta = 60^\circ$, for
different values of the energy, $E_\nu \in [5, 50]$ GeV. The effect of
the different sign in front of the $Y$-term in
eqs.~(\ref{eq:spagnoli}) and (\ref{eq:etau}) can be seen in the
opposite shape in the ($\theta_{13}, \delta$) plane of the $\nu_e \to
\nu_\tau$ curves with respect to the $\nu_e \to \nu_\mu$ ones.  Notice
that both families of curves meet in the ``physical'' point,
$\theta_{13} = \bar \theta_{13}$, $\delta = \bar \delta$, and any
given couple of curves belonging to the same family intersect in a
second point that lies in a restricted area of the ($\Delta \theta,
\delta$) plane, the specific location of this region depending on the
input parameters ($\bar \theta_{13}, \bar \delta$) and on the neutrino 
energy. Using a single set of experimental data (e.g. the golden muons), 
a $\chi^2$ analysis will therefore identify two allowed regions: the ``physical''
one (around the physical value, $\bar \theta_{13}, \bar \delta$) and
a ``clone'' solution, spanning all the area where a second
intersection between any two curves occurs.  This is the source of the
intrinsic ambiguity pointed out in~\cite{Burguet-Castell:2001ez}. When
considering at the same time experimental data coming from both the
golden and the silver muons, however, we see that the
clone regions for each set of data lie well apart. Thus, a comprehensive
$\chi^2$ analysis of the data will in principle result in
the low-$\chi^2$ region around the physical pair, only.  Of course,
this statement is only true if the statistical significance of both
sets of experimental data is sufficiently high.

\section{Expected rates at the Neutrino Factory}
\label{sec:rates}

\begin{figure}[h!]
\begin{center}
\epsfig{file=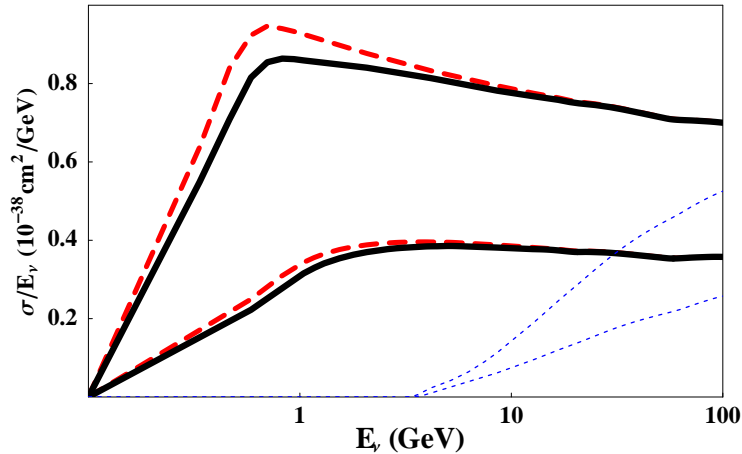,width=10cm}
\end{center}
\vspace{-0.5cm}
\caption{\label{fig:crosst} \it Behaviour of the (anti)neutrino-nucleon cross
section as a function of the neutrino energy for $\nu_e (\bar \nu_e)$ 
(long dashed lines), $\nu_\mu (\bar \nu_\mu)$
(continuum lines) and $\nu_\tau (\bar \nu_\tau)$ (dashed lines) 
{\rm~\cite{Paolo}}.}
\end{figure}

To get the expected number of events per bin, we must now convolute
the neutrino fluxes at the Neutrino Factory with the charged-current (CC)
cross-section and with the transition probability in
eqs.~(\ref{eq:spagnoli}) and (\ref{eq:etau}). At low energies the
neutrino scattering cross-section is dominated by quasi-elastic
scattering and resonance production (see~\cite{Lipari:2002at,Zardetto}
and references therein).  However, if $E_\nu$ is greater than
$\sim10$~GeV, which is the case for high-energy muons in the storage
ring (30-50 GeV), the total cross-section is dominated by deep
inelastic scattering (DIS) and can be approximately described with the
cross-section on an isoscalar target~\cite{Conrad:1997ne} : \bea
\sigma(\nu +N \to \ell^- + X) &\approx& 0.67\times 10^{-42} \; \times
\frac{E_{\nu}}{GeV} \times m^2 \nn \, , \\ & & \label{isoscalar}\\
\sigma(\bar{\nu} +N \to \ell^+ + X) &\approx& 0.34\times10^{-42} \;
\times \frac{E_{\bar \nu}}{GeV} \times m^2 \nn.  \eea
Eq.~(\ref{isoscalar}) is not a good description of the DIS
cross-section for $\nu_\tau$ interactions in the considered range of
energies, since the lepton mass cannot be neglected. Taking into
account these effects, the total cross-section for the $\nu_x (\bar
\nu_x)-N$ interactions appears as in Fig.~\ref{fig:crosst}, where we
show its behaviour as a function of the neutrino energy.
The theoretical number of expected charged-current events 
(with production of final leptons $\ell_\beta$) in the detector, for a fixed
parent muon energy $E_\mu$, is: 
\bea
N_{\ell_\beta}(\nu_\alpha \to \nu_\beta)& = & N_{Kton}\times N_{A}\times 10^{9}
\int_{E_\nu} d E_\nu \, \sigma_{\nu_\beta} (E_\nu) 
\,  P_{\alpha \beta} (E_\nu; \bar \theta_{13}, \bar \delta)
\, \frac{d\Phi_\alpha(E_\nu)}{d E_\nu} 
\label{nu_ev}
\eea where $N_{Kton}$ is the detector mass in Kton and $N_{A}\times 10^{9}$ 
is the number of nucleons per Kton. 

In Tab.~\ref{tab:tab3} we report the neutrino and anti-neutrino
charged-currents interaction rates assuming no oscillation for two
different baselines (namely $L$ = 732 and 3000 km) for an ideal
(i.e. with perfect efficiency) detector with a mass of 1 Kton. We
consider three different muon polarization values
$\cal{P}_{\mu^{\mp}}$=0, $\pm 0.3$ (the "natural" polarization) and
$\pm 1$.  The neutrino beam results from the decay of $ 2 \times
10^{20} \mu^+$'s and $\mu^-$'s per year in a straight section of an
$E_\mu = 50$ GeV muon accumulator, with five operational years
for each polarity. Fluxes have been integrated in the forward
direction with an angular divergence (taken to be constant) $\delta
\varphi \sim 0.1$ mr. Beam divergence effects and QED one-loop radiative 
corrections to the neutrino fluxes have been taken into account 
in~\cite{Broncano:2002hs}. The overall correction
to the neutrino flux has been shown to be very small, ${\cal O} (0.1 \%)$.

The situation in the hypothesis of neutrino oscillations at the same
baselines is given in Tabs.~\ref{tab:lept732mm} and
\ref{tab:lept3000mm}. The considered PMNS matrix parameters are:
$\theta_{12} = 33^\circ$, $\Delta m_{12}^2=10^{-4}$ eV$^2$;
$\theta_{23}=45^\circ$, $\Delta m_{23}^2=2.9 \times 10^{-3}$ eV$^2$;
$\theta_{13} = 5^\circ$ and $\delta = 90^\circ$.

\begin{table}[h!]
\centering
\begin{tabular}{||c|c||c|c||c|c||}
\hline \hline
\multicolumn{2}{||c||}{$E_{\mu^{\mp}}= 50$ GeV} & 
\multicolumn{2}{c||} {$\mu^{-}$} &  
\multicolumn{2}{c||}{$\mu^{+}$}\\
\hline
\multicolumn{2}{||c||}{L (km)} & $N_{\nu_\mu}/10^4$ & $N_{\bar\nu_e}/10^4$  & 
$N_{\bar\nu_\mu}/10^4$ &$N_{\nu_e}/10^4$ \\  
\hline\hline
\multicolumn{2}{||c||}{732}& \multicolumn{2}{r} {} & 
\multicolumn{2}{l||} {} \\
\hline\hline
   & $0$       & $173$ & $75$ & $88.1$ & $148$\\
\cline{2-6}
$\cal{P}_{\mu^{\mp}}$
   & ${\pm}0.3$& $151$ & $97.5$ & $76.7$ & $192$\\
\cline{2-6}
   & ${\pm}1$  & $98.4$ & $150$ & $50$ & $295$\\ 
\hline\hline
\multicolumn{2}{||c||}{3000} & \multicolumn{2}{r} {} & 
\multicolumn{2}{l||} {} \\
\hline\hline
   & $0$       & $10.3$ & $4.47$ & $5.24$ & $8.79$\\
\cline{2-6}
$\cal{P}_{\mu^{\mp}}$
   & ${\pm}0.3$& $8.98$ & $5.81$ & $4.56$ & $11.4$\\
\cline{2-6}
   & ${\pm}1$  & $5.86$ & $8.93$ & $2.98$ & $17.6$\\ 
\hline\hline
\end{tabular}
\caption{\it{ Neutrino and anti-neutrino charged-currents interaction
rates for $L$ = 732~km and 3000~km per 1 Kton and per 5 operational
years when $2 \times 10^{20}$ muons decay in the straight section of
the storage ring. These fluxes have been averaged over an angular
divergence of 0.1 mr. The results are presented for three muon
polarizations $\cal{P}_{\mu^{\mp}}$$=0, \pm 0.3 \,and \,\pm 1$.}}
\label{tab:tab3}
\end{table}

\begin{table}[h!]
\centering
\begin{tabular}{|c|c|c|c|c|c|c|}
\hline
{\phantom{\Huge{l}}}\raisebox{-.2cm}{\phantom{\Huge{j}}}
$\boldmath \cal{P}_{\mu^{-}}$ &  $N_{\mu^-}/10^4$ &$N_{e^+}/10^4$&
$N_{\mu^+}$ & $N_{e^-}$ & $N_{\tau^+}$ & $N_{\tau^-}/10^2$\\
\hline
{\phantom{\Huge{l}}}\raisebox{-.2cm}{\phantom{\Huge{j}}}
$0$  & 172 & 75 & 107 & 186 & 80.7 & 89.9 \\
{\phantom{\Huge{l}}}\raisebox{-.2cm}{\phantom{\Huge{j}}} 
 $0.3$     & 150 & 97.5 & 140 & 174 & 105 & 81.7 \\
{\phantom{\Huge{l}}}\raisebox{-.2cm}{\phantom{\Huge{j}}} 
$1$      & 97.5 & 150 & 215 & 147 & 161 & 64.6 \\
\hline
{\phantom{\Huge{l}}}\raisebox{-.2cm}{\phantom{\Huge{j}}} 
$\cal{P}_{\mu^{+}}$ &  $N_{\mu^+}/10^4$ &$N_{e^-}/10^4$&
$N_{\mu^-}$ & $N_{e^+}$ & $N_{\tau^-}$ & $N_{\tau^+}/10^2$\\
\hline
{\phantom{\Huge{l}}}\raisebox{-.2cm}{\phantom{\Huge{j}}}
$0$  & 87.4 & 148 & 244 & 99 & 151 & 45.2 \\
{\phantom{\Huge{l}}}\raisebox{-.2cm}{\phantom{\Huge{j}}} 
 $-0.3$     & 76.1 & 192 & 317 & 93.4 & 196 & 41.5 \\
{\phantom{\Huge{l}}}\raisebox{-.2cm}{\phantom{\Huge{j}}}
$-1$      & 49.5 & 295 & 487 & 79.3 & 302 & 32.8 \\
\hline
\end{tabular}
\caption{\it  
Oscillated charged-currents event rates for $\mu^-$ (upper table) and $\mu^+$
(lower table) beam assuming neutrino oscillations with $\theta_{13} = 5^\circ$
and $\delta = 90^\circ$ in a 1 Kton detector, for a L = 732 km baseline for different 
polarizations of the parent muon. We have considered $1 \times 10^{21}$ 
muons decays ($2 \times 10^{20}$ useful muons/year $\times$ 5 operational 
years).}
\label{tab:lept732mm}
\end{table}

\begin{table}[h!]
\centering
\begin{tabular}{|c|c|c|c|c|c|c|}
\hline
{\phantom{\Huge{i}}}\raisebox{-.2cm}{\phantom{\Huge{j}}} 
$\cal{P}_{\mu^{-}}$ &  $N_{\mu^-}/10^3$ &$N_{e^+}/10^3$&
$N_{\mu^+}$ & $N_{e^-}$ & $N_{\tau^+}$ & $N_{\tau^-}/10^2$\\
\hline
{\phantom{\Huge{l}}}\raisebox{-.2cm}{\phantom{\Huge{j}}}
$0$  & 92.2 & 44.6 & 48.7 & 154 & 51.5 & 83.3 \\
{\phantom{\Huge{l}}}\raisebox{-.2cm}{\phantom{\Huge{j}}}
$0.3$     & 79.6 & 58 & 63.2 & 143 & 67 & 76 \\
{\phantom{\Huge{l}}}\raisebox{-.2cm}{\phantom{\Huge{j}}} 
$1$      & 50.2 & 89.2 & 97.3 & 120 & 103 & 59 \\
\hline
{\phantom{\Huge{l}}}\raisebox{-.2cm}{\phantom{\Huge{j}}} 
$\cal{P}_{\mu^{+}}$ &  $N_{\mu^+}/10^3$ &$N_{e^-}/10^3$&
$N_{\mu^-}$ & $N_{e^+}$ & $N_{\tau^-}$ & $N_{\tau^+}/10^2$\\
\hline
{\phantom{\Huge{l}}}\raisebox{-.2cm}{\phantom{\Huge{j}}}
$0$  & 46.8 & 87.5 & 245 & 63.2 & 126 & 42.4 \\
{\phantom{\Huge{l}}}\raisebox{-.2cm}{\phantom{\Huge{j}}} 
$-0.3$     & 40.5 & 114 & 319 & 58.4 & 164 & 38.7 \\
{\phantom{\Huge{l}}}\raisebox{-.2cm}{\phantom{\Huge{j}}}  
$-1$      & 25.6 & 175 & 491 & 47 & 252 & 30 \\
\hline 
\end{tabular}
\caption{\it  The same as Tab.~\ref{tab:lept732mm} but for $L=3000$ km.}
\label{tab:lept3000mm}
\end{table}


\section{Detector issues}
\label{sec:issues}
In this Section we discuss the basic performance of an OPERA-like
detector at the Neutrino Factory in reconstructing neutrino
interactions. In doing so, we profit of the experience and of the
test results gathered within the OPERA 
Collaboration~\cite{proposal,Guler:2001hz}.

\subsection{The detector layout }
 
The main features of the detector we are going to discuss in this
paper are very similar to the ones of the OPERA
experiment~\cite{proposal,Guler:2001hz}, which is meant for a long
baseline search for $\nu_{\mu}\rightarrow\nu_{\tau}$ oscillations at
the CNGS beam.  The experiment uses nuclear emulsions as high
resolution tracking devices for the direct detection of the $\tau$
produced in the $\nu_{\tau}$ CC interactions with the target.

\begin{figure}[h!]
\begin{center}
\epsfig{file=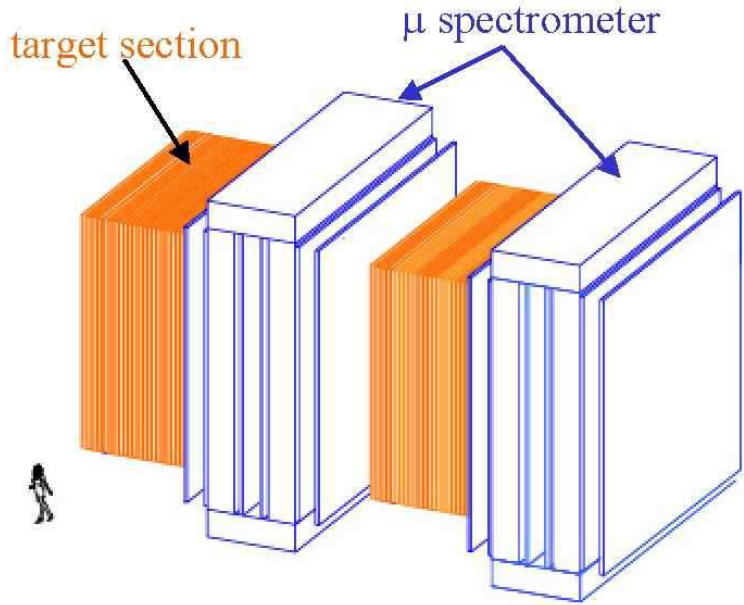,width=10cm}
\end{center}
\vspace{-0.5cm}
\caption{\label{fig:opera_cngs}  \it OPERA at CNGS.}
\end{figure}

OPERA is designed starting from the Emulsion Cloud Chamber concept
(see references quoted in~\cite{proposal}) which combines the high
precision tracking capabilities of nuclear emulsions and the large
mass achievable by employing metal plates as a target. The basic
element of the ECC is the cell which is made of a 1~mm thick lead
plate followed by a thin emulsion film.  The film is made up of a pair
of emulsion layers $50~\mu$m thick on either side of a $200~\mu$m
plastic base.  Charged particles give two track segments in each
emulsion film.  The number of grain hits in about $50~\mu$m
($15$-$20$) ensures redundancy in the measurement of particle
trajectories. By piling-up a series of cells in a sandwich-like
structure bricks can be built, which constitute the detector element
for the assembly of massive planar structures (walls). A wall and
its related electronic tracker planes constitute a module. A
supermodule is made of a target section, which is a sequence of
modules, and of a downstream muon spectrometer. The detector consists
of a sequence of supermodules (see Fig.~\ref{fig:opera_cngs}).

The signal of the occurrence of a $\nu_\tau$CC interaction in the
detector target is identified by the detection of the $\tau$ lepton in
the final state through the direct observation of its decay
topology. A $\tau$ may decay either into the lead plate where the
interaction occurred (short decay) or further downstream (long
decay). For long decays, the $\tau$ is detected by measuring the angle
between the charged decay daughter and the parent direction. The
directions of the tracks before and after the kink are reconstructed
in space by means of the pair of emulsion films sandwiching the lead
plate where the interaction occurred. A fraction of the short decays
is detectable by measuring a significant impact parameter (IP) of the
daughter track with respect to the tracks originating from the primary
vertex.

The detection of the $\tau$ decay and the background reduction benefit
from the dense brick structure given by the ECC, which allows the
electron identification through its showering, the pion and muon
separation by the $dE/dx$ measurement method, and the determination of
the momentum of each charged particle employing techniques based on
the Multiple Coulomb Scattering. All these methods are discussed in
the following.

Electronic detectors placed downstream of each emulsion brick wall are
used to select the brick where the interaction took place (to be removed for 
the analysis) and to guide the emulsion scanning. The target
electronic detectors are also used to sample the energy of hadronic
showers and to reconstruct and identify penetrating tracks.

In the following we assume a supermodule structure with 31 brick walls
followed by a muon spectrometer. For the present analysis, we consider
the same setup as for OPERA at CNGS. The electronic detectors are 1~cm
thick, 2.6~cm wide scintillator bars. The active detectors at the
spectrometers are Resistive Plate Chambers (RPC) with 3~cm wide
readout strips. The bending of the particle trajectory before and
after the magnet is measured by high precision drift tubes
(DT)~\cite{proposal}.  With this configuration the total target mass
of a supermodule is 0.9 Kton.

For details on the event reconstruction both with the nuclear
emulsions and the electronic detector, and on the OPERA sensitivity to
$\nu_{\mu}\rightarrow\nu_{\tau}$ and $\nu_{\mu}\rightarrow\nu_{e}$
oscillations, we refer to~\cite{proposal,komatsu,Guler:2001hz}.

\subsection{Kinematical analysis with ECC }

\subsubsection{Momentum measurement through the multiple scattering}

When a particle of momentum $p$ and velocity $\beta$ traverses a
material of thickness $X$, measured in units of radiation length
$X_0$, the distribution of the scattering angle $\theta_0$ in a plane is
approximately gaussian with a RMS given by
\begin{equation}
\theta_0 = \frac{13.6~\mbox{MeV}}{p\beta} \sqrt{X}\, .
\end{equation} 

\begin{figure}[tb]
  \begin{center}
    \resizebox{0.8\textwidth}{!}{
      \includegraphics{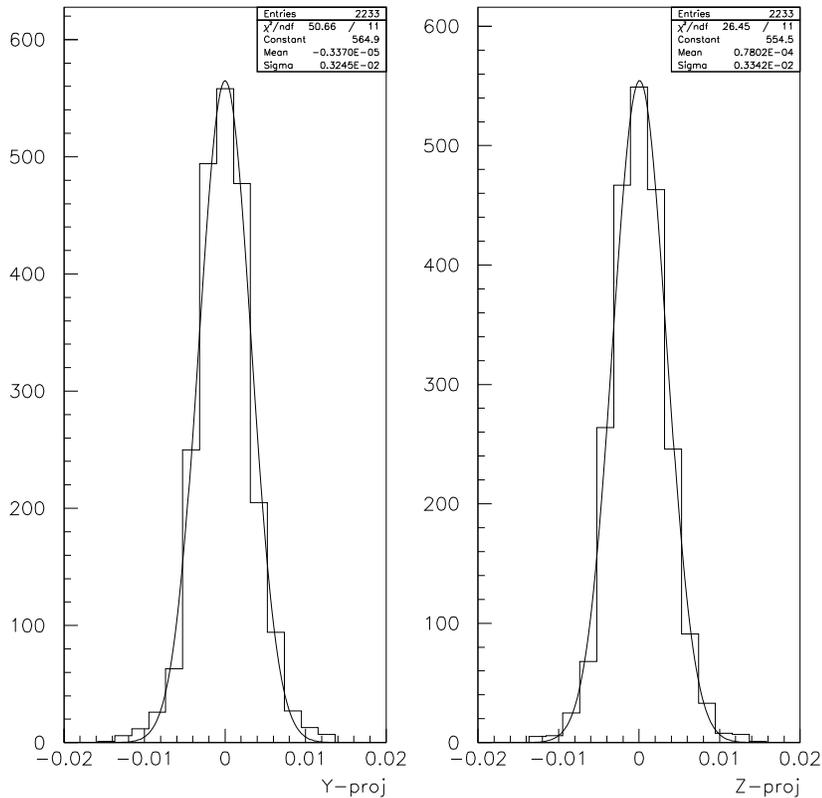}
}
      \caption{\it The measured angular resolution routinely achieved
with an ECC shown for both angular projections.
        \label{fig:resolution}}
  \end{center}
\end{figure}

One can determine $\theta_0$ by measuring the distribution of the
angular difference in two consecutive emulsion films. This method is
called {\em angular method}. A resolution $\delta \theta$ of about 2
to 3 mrad can be routinely obtained as shown in
Fig.~\ref{fig:resolution}. The angular resolution is affected by both
systematical sources (i.e. planarity of the scanning surface) and
statistical effects which can be significantly reduced by means of
precise and repeated measurements respectively providing a resolution
of about 1 mrad.

The RMS of the measured scattering angles, $\theta^{2}_{M}$, is the 
quadratic sum of the scattering signal $\theta_0$ and of the measurement 
error $\delta \theta$: $\theta^{2}_M = \theta^{2}_0 + \delta \theta^2$. 
The error on $\theta_0$ when $N$ 
independent measurements are performed is 

\begin{equation}
\frac{\delta \theta_0}{\theta_0} = \frac{1}{\sqrt{1-(\delta 
\theta/\theta_M )^2}} \frac{1}{\sqrt{N}}. 
\end{equation}

\begin{figure}[tb]
  \begin{center} \resizebox{0.8\textwidth}{!}{
    \includegraphics{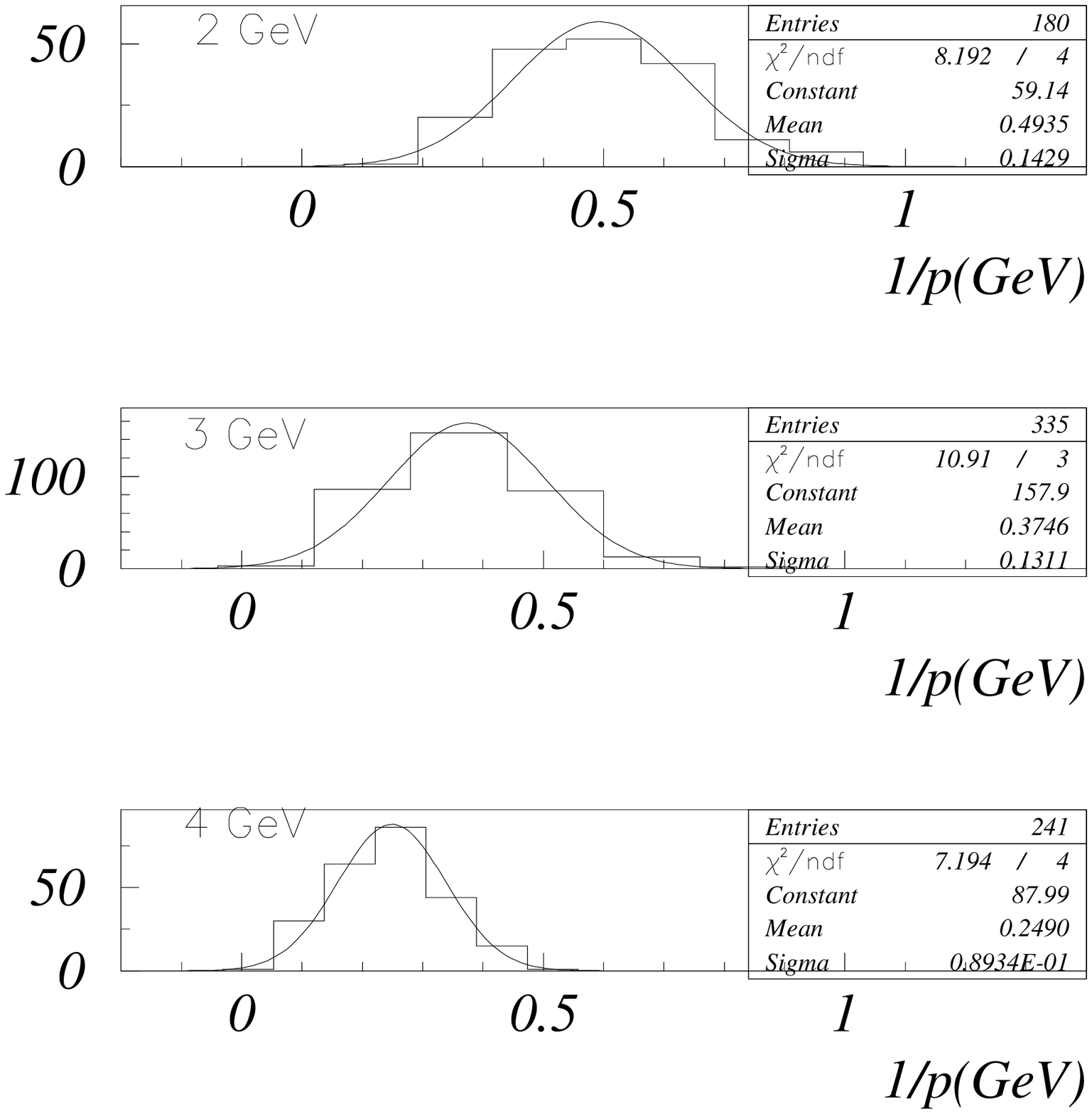}} \caption{\it The $1/p$
    resolution achieved with a 3$X_0$ ECC and 3~mrad angular
    resolution exposed to 2, 3 and 4~GeV pions.  \label{fig:mcsang}}
    \end{center}
\end{figure}

Using $N_{film}$ emulsion films, $N = (N_{film}-1) \times 2$ points
can be used, since each of the two projections provides a set of
independent measurements. Therefore, one can easily obtain a $\delta
p/p \sim 10\%$ with a brick of about 50 emulsion films. In
Fig.~\ref{fig:mcsang} we report the momentum resolution corresponding
to an angular resolution of 3 mrad for 2, 3 and 4 GeV negative pions
impinging onto a 3X$_0$ ECC (17 emulsion films)~\cite{mcsnim}. A
resolution of 28\% (2 GeV) to 36\% (4 GeV) has been achieved.

\subsubsection{Electron identification}
\begin{figure}[tb]
  \begin{center}
    \resizebox{0.8\textwidth}{!}{
      \includegraphics{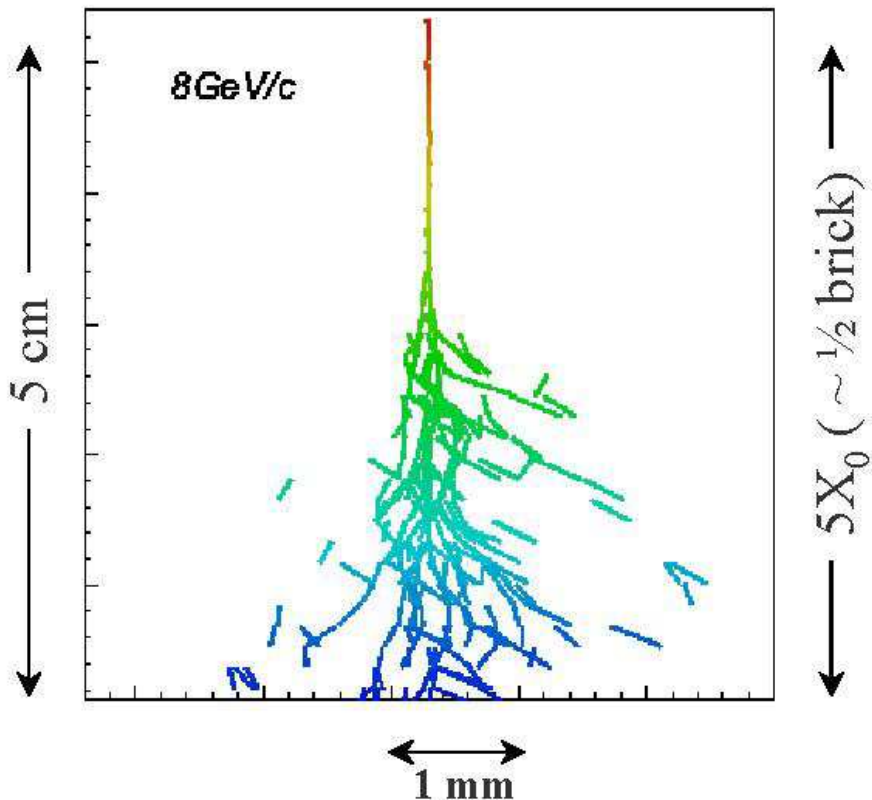}}
      \caption{\it Electromagnetic shower observed in a brick exposed to an 
electron beam. 
        \label{fig:prop121}}
  \end{center}
\end{figure}

A typical example of an electron identified in an ECC brick is shown
in Fig.~\ref{fig:prop121}. The identification is done employing the
multiple scattering technique discussed above applied to the electron
track before showering and counting the number of tracks when the
shower develops (calorimetric measurement). In particular the peculiar
energy loss rate of electrons is used to discriminate between
electrons and hadrons: the energy is practically unchanged for hadrons
but it is sizebly decreased by bremsstrahlung for electrons while
traversing the brick. A $\chi^2$ estimator to test the two possible
hypotheses (hadrons or electrons) is built and it is used as a
separator. The $\chi^2$ minimization provides also the particle energy
estimation. The other way to identify electrons is to count segments
associated to its shower. The lateral spread of a shower is of the
order of 1~cm which makes the shower itself well confined within a
single brick. An energy resolution of about 20\% can be obtained with
this method, limited by background tracks falling into the analysis
region.

Preliminary studies show that combining the two methods it is possible
to achieve 97\% identification efficiency for electrons with energy
above 1~GeV~\cite{proposal}.

\subsubsection{Low momentum muon identification}
\label{sec:low_momentum}

Low momentum muons stopping in the ECC can be identified measuring the
energy loss rate near the end-point of the range~\cite{powell}. In
this regime, the energy loss rate is given by $I = dE/dx =
k/\beta^2$. A measurement of the energy loss rate $I$ and of the distance 
$x$ traveled by the particle before stopping will therefore give the 
particle mass $M = (2/\beta^2) E = (2/k) I^2 x$, with an error that
gets contributions from $\delta x$ and $\delta I$. The latter is the 
dominant one when a particle stops in the brick. 

The grain measurement amounts to about $(30\pm\sqrt{30})/100~\mu$m; hence, 
by measuring only one emulsion sheet one can get about 18\% resolution 
on the ionization measurement. Since muons and pions are expected to cross 
about 10 emulsion films after entering into the non-relativistic regime, 
a mass resolution $\delta M/M \simeq 2\delta I/I = 2 \times (0.18 / \sqrt{10}) 
\simeq 0.12$ is achievable which means $\delta M = 16$~MeV for a pion. 
This mass resolution is used to separate muons from pions.
 
\subsection{Muon reconstruction }
\label{muonsec}

The magnetic spectrometers located in the back of the ECC target are
aimed to reconstruct the charge and the momentum of penetrating
particles.  The fraction of ``silver'' candidates where it is possible
to measure the muon sign and the corresponding mis-assignment
probability is a key parameter to assess the performance of OPERA as a
detector for the Neutrino Factories.  In fact the primary muon
identification, even without charge reconstruction, plays an important
role to veto background events. In particular, charm-production from
$\bar \nu_\mu$~CC interactions can be significantly suppressed if a
prompt muon from the primary vertex is identified.  Both at CNGS and
in the present analysis muon candidates are selected combining the
information of the scintillator trackers and the spectrometer RPC's
through a pattern recognition algorithm.  For a detailed description
of this procedure and the build-up of the candidate tracks we refer
to~\cite{proposal} (Section~5.4) and~\cite{Guler:2001hz}
(Section~5.3).  The muon identification algorithm is described in
details in~\cite{proposal} (Section.~7.2) and~\cite{Guler:2001hz}
(Section~5.4) and it is briefly recalled here.

\subsubsection{Muon identification}
Among the reconstructed tracks, the longest one is chosen as a
possible muon candidate. If the track is very penetrating and exits
from the back of the spectrometer, it is validated as muon and no
further cuts are applied. Otherwise cuts are imposed on the number of
target walls and spectrometer iron plates belonging to the candidate
and on the isolation of the track.  Finally, a candidate is validated
if the momentum measured by range is consistent (within proper
tolerances) to the one measured in the ECC through Multiple Coulomb
Scattering.  All the ECC tracks matching the muon candidate within
200~mrad are considered. If more than one ECC track exhibits
range-momentum correlation, the one with the best angular match is
selected.  The performance of the algorithm can be expressed in term
of muon identification efficiency and matching probability.  The
former represents the number of true muons passing the $\mu$-id cuts
and matched to a track in the emulsions. The latter quantifies the
probability of correct matching to ECC.
Fig.~\ref{fig:muons_from_taumu} shows the muon identification
efficiency and the matching probability versus the energy of the muon
in $\nu_\tau \rightarrow \tau^- X \rightarrow \mu^- \bar \nu_\mu
\nu_\tau X$ interactions.  It is worth noting that the muon
identification efficiency for short muons stopping before the
spectrometer can be improved significantly measuring the $dE/dx$ of
the particles in the emulsions around the range-out area (see
Section~\ref{sec:low_momentum}). This is particularly interesting for
charm vetoing and in the OPERA experiment running at the CNGS allows a
reduction of the charm background by about a factor 2. However, these
results are only based on Monte Carlo simulations and tests are in
progress in order to measure experimentally the $\pi/\mu$ separation
through the $dE/dx$ technique. Although preliminary results are in a
good agreement with Monte Carlo estimates, conservatively for the
present analysis this technique is not employed. However, as an
example, we will give results by using this technique only for a given
($\bar{\theta}_{13},\bar{\delta}$) pair, see
Section~\ref{sec:o732i3000} .



\subsubsection{Charge measurement}

\begin{figure}[htbp]
\begin{center}
\begin{tabular}{c}
\hspace{-0.3cm}
\epsfxsize12.5cm\epsffile{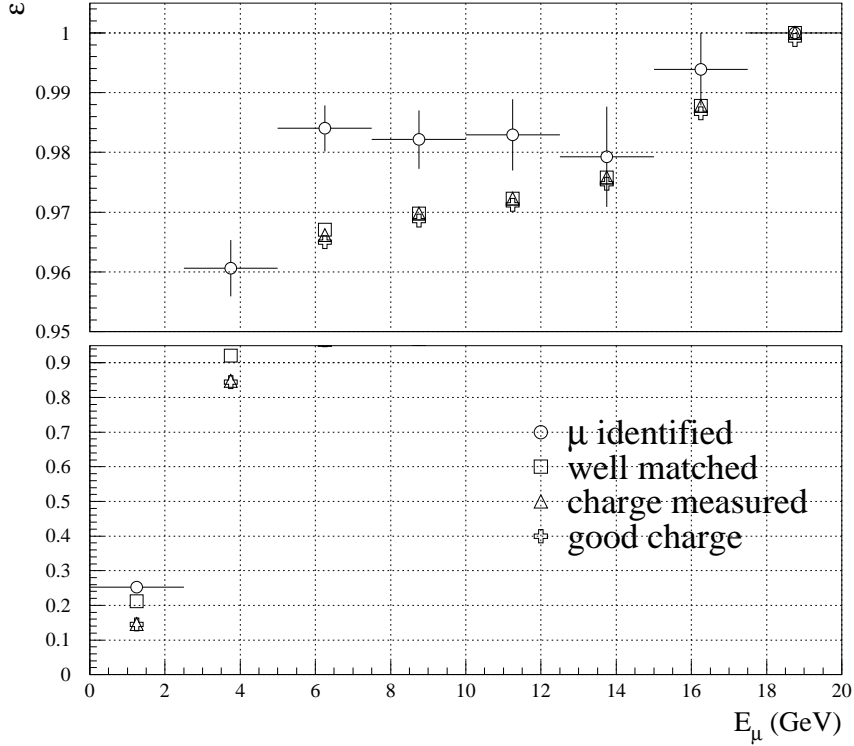}
\end{tabular}
\caption{\it Percentage of 
$\nu_\tau \rightarrow \tau^- X 
\rightarrow \mu^- \bar \nu_\mu \nu_\tau X$ 
events with muon identified (dots), well
matched to the ECC track (square), reaching the spectrometer (triangle)
and with charge correctly assigned (crosses)
versus the energy of the muon. To ease the reading,
the width of the bins and the statistical errors are shown just for the
first mark.}
\label{fig:muons_from_taumu}
\end{center}
\end{figure}

OPERA is able to measure the charge of the muons crossing the magnetic
spectrometers or stopping in them. For through going muons an iterative
$\chi^2$ minimization procedure is used to fit the measured muon
trajectory combining the hits from target scintillators, spectrometer
RPC's and drift tubes. Hence, both the particle momentum and the
charge can be reconstructed. For stopping particles the best estimate
of the momentum is provided by the range, while the charge
reconstruction still relies on the curvature in the magnetic
field. The fraction of events where the charge can be measured
(through going muons and stopping muons with enough spectrometer hits)
and the probability to assign correctly the charge are shown in
Fig.~\ref{fig:muons_from_taumu} as a function of the muon energy. As
already discussed in~\cite{proposal} the charge mis-assignment
probability in the energy range of interest is approximately constant
and it is at the level of 0.1-0.3\%.

\section{Analysis of silver muons }
\label{sec:anasilver}
Silver wrong-sign muons are produced in $\nu_\tau$CC interactions,
coming from $\nu_e$ oscillations with subsequent muonic decay of the
$\tau$. In the following we discuss in detail the signal efficiency
and the expected background under the hypothesis that $\mu^+$'s
circulate into the muon storage ring (namely, a 50\% $\nu_e$ plus 50\%
$\bar \nu_\mu$ beam hits the detector).  Therefore, the signal we are
looking for is
$\nu_e\rightarrow\nu_\tau\rightarrow\tau^-\rightarrow\mu^-$.

\subsection{Expected number of signal events}
\label{expesigna}

The search for muonic tau decays is performed by using an approach
very similar to the one described in Refs.~\cite{proposal,Guler:2001hz}.  
Two classes of events are considered: {\em short} (the $\tau$ decays into 
the lead plate where the interaction occurred) and {\em long} decays 
(the decay occurrs outside the vertex plate, therefore the kink angle 
in space can be reconstructed). 

\begin{figure}[htbp]
\begin{center}
\begin{tabular}{c}
\hspace{-0.3cm}
\epsfxsize12.5cm\epsffile{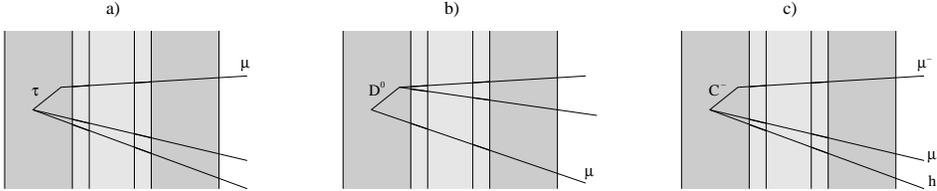}
\end{tabular}
\caption{\it Sketch of signal a) and background b) and c) topologies
for $\tau\rightarrow\mu$ decays.}
\label{fig:shortback}
\end{center}
\end{figure} 

The short decay search exploits the impact parameter technique,
searching for a muon with a large IP with respect to a reconstructed
vertex of at least two tracks (see Fig.~\ref{fig:shortback}). The main
background is a subset of charm-production. Charm can be produced in
$\bar{\nu}_\mu$~CC and contributes to the signal either when the
charge of the primary muon is wrongly measured in case of neutral
charmed-hadron production (Fig.~\ref{fig:shortback}-b) or when the
primary muon is not identified in the case of charged charmed-hadron
production (Fig.~\ref{fig:shortback}-c).  Charm can also be produced
in ${\nu}_\mu$ (from $\nu_e\rightarrow\nu_\mu$) CC interactions. If
the origin of the muon track is unknown or wrongly reconstructed,
these events can mimick a $\tau\rightarrow\mu$ decay (as sketched in
Fig.~\ref{fig:shortback}).

This background is significantly reduced by imposing a lower cut at
2~GeV on the invariant mass of the hadronic system, as computed by the
momenta measured in the ECC. Even with a modest momentum resolution of
50\% the charm background can be reduced by more than a factor of 1000
retaining 15\% of the $\tau\rightarrow\mu$ short decay signal. The
results reported in~\cite{Guler:2001hz} show that the total
efficiency for muonic short decays is about 0.7\%, including the
muonic branching ratio of the $\tau$. Although this analysis is very
preliminary and could be improved, in the following we assume such an
efficiency. The expected background was evaluated for the CNGS
conditions to be $2\times10^{-6}\times\nu_\mu\mbox{CC}$ DIS events,
which is very small. Moreover, having in mind that at a Neutrino
Factory the bulk of the events comes from $\bar{\nu}_\mu$
interactions, while at the CNGS from ${\nu}_\mu$, we can safely assume
that the short decay channel is background free.

For long decays, $\tau$ candidates are searched for by measuring the
kink angle in space and taking into account the worse angular
resolution for decays in the base. The inefficiency in the kink
detection results from the need of rejecting small-angle and very
large-angle kinks. The upper cut ($>$ 500~mrad) is motivated by
considerations related to the scanning efficiency of the automatic
microscopes. The lower cut (20~mrad in space) has been assigned taking
into account that experimental data for hadron rescattering are
available only down to such an angle. The excellent intrinsic angular
resolution of the emulsion films would allow a substantially looser
cut. Nevertheless, awaiting background measurements extended to lower
angles, we maintain the conservative 20~mrad cut. As in the analysis
described in Ref.~\cite{proposal} we only consider decays occurring in
two lead plates downstream from the vertex plate. Unlike the CNGS
analysis described in the OPERA Proposal, we do not apply neither a
cut on the muon momentum nor a cut on the visible energy. On the other
hand the charge measurement of the muon track is required. The only
kinematical cut we apply is the transverse momentum at the decay
vertex to be larger than 250~MeV. This cut allows the rejection of
muonic decays of pions and kaons.

Summing up the short and long decay contributions, the overall
efficiency (including the muonic tau decay branching ratio) for an
average neutrino energy of 35~GeV is about 5\%.

The expected number of signal events detected during a 5 year data
taking with a 1~Kton detector for different values of
$(\theta_{13},\delta)$ is given in Tab.~\ref{tab:sigeve}.

\begin{table}[hbtp]
  \small
\begin{center}
\vspace{5mm}
\begin{tabular}{||c|c|c||}
\hline
\hline
$(\theta_{13},\delta)$ & $\tau^-\rightarrow\mu^-$  & $\tau^-\rightarrow\mu^-$\\
                       &  L = 732~km       &  L = 3000~km    \\
\hline
$(1^\circ,-90^\circ)$ & 0.64 & 0.72 \\
\hline
$(5^\circ,-90^\circ)$ & 8.73 & 8.90 \\
\hline
$(1^\circ,0^\circ)$ & 0.0092 & 0.070\\
\hline
$(5^\circ,0^\circ)$ & 5.57 & 5.66 \\
\hline
$(1^\circ,90^\circ)$ & 0.52 & 0.30 \\
\hline
$(5^\circ,90^\circ)$ & 8.11 & 6.81 \\
\hline
\hline
\end{tabular}
\caption{\it Expected number of muonic tau decays detected into a
1~Kton OPERA-like detector for various $(\theta_{13},\delta)$ values
and baselines (i.e., the $\tau^\pm \to \mu^\pm$ branching ratio and
the overall estimated efficiency are taken into account).  We have
considered $1 \times 10^{21}$ muons decays with ``natural''
polarization ($2 \times 10^{20}$ useful muons/year $\times$ 5
operational years) for each polarity.}
\label{tab:sigeve}
\end{center}
\end{table}

\subsubsection{The golden muons signal at an ECC detector }
\label{expegolden}


Although not competitive with large magnetized calorimeters, an
OPERA-like detector can also be used to study
$\nu_e\rightarrow\nu_\mu$ oscillations. From an experimental point of
view this search does not imply additional efforts with respect to the
ones needed to study the silver channel. Indeed, as soon as a wrong
sign muon is reconstructed in the detector the brick where the
interaction occurred is removed and the event carefully analyzed. If a
decay candidate is found the event falls in the silver channel sample
and the appropriate kinematical analysis applied. Conversely, it is
classified as a golden event.



In an OPERA-like detector the background from charmed hadron and
$\tau$ (produced by $\nu_\tau$ coming from $\nu_e\rightarrow\nu_\tau$
oscillations) decays is highly suppressed because of the detector
capability in detecting decay topologies. Since the presence of a kink
on the muon track is not required, the background from $h^-$ decays or
$h^-$ punch-through in this channel is higher.

Given the fact that, also in the most optimistic case, the mass of the
OPERA-like detector is small if compared to a MID detector and it is
not fully magnetized, the gain in sensitivity one could have in
studying this channel is modest. Therefore, in the following we focus
only on the $\nu_e\rightarrow\nu_\tau$ sensitivity and we use it in
combination with a MID detector exploiting the golden channel.





\subsection{Expected background in the long decay sample}

\subsubsection{Neutrino induced charm-production}
\label{sec:nu_induced_charm}
Charm-production from neutrinos can be induced by both 
unoscillated $\nu_e$ and $\nu_\mu$ coming from oscillated
$\nu_e$ through the reaction

$$ \nu_l N \rightarrow l^- + C^+ + X\, $$

If the primary lepton is not identified and the charge of the muon
from the semi-leptonic decay of the charmed hadron is wrongly
measured, then the event is classified as a silver candidate.

The charm-production rate has been estimated by using the results of a
combined analysis of all available data on neutrino-induced
charm-production~\cite{DeLellis:2002mf}. By using the $\nu_\mu,\,
\nu_e$ predicted spectrum, we expect a charm rate, normalized to CC
interactions, of

$$R_c(\nu_\mu) = (5.12\pm0.30)\%;\, R_c(\nu_e) = (4.74\pm0.23)\%\, .$$

Given the capability in detecting decay topologies, an ECC detector is
only sensitive to semi-muonic decays of the charged charmed mesons,
unlike the electronic detectors. As it can be derived from
Ref.~\cite{fractions}, the fraction of charged among produced charmed
particles is, at the Neutrino Factory energies, of about 45\%
($f_{C^+})$. Moreover, given the fact that the search is limited to
muonic tau decays, we have to consider only the semi-muonic branching
ratio of charmed hadrons, which amounts to about
10\%~\cite{Kayis-Topaksu:2002xq}.

Finally, we can write the expected number of events from this background as

$$N_c = N_{l}^{CC}\times R_c(l)\times
f_{C+}\times(1-\varepsilon_{lID})\times BR(C^+\rightarrow
l^+)\times\varepsilon_{\mu^+ID}\times(1-\varepsilon_{charge})\times\varepsilon_{det}$$

where $N_{l}^{CC}$ is the total number of charged-current events
induced by $\nu_l$, $\varepsilon_{lID}$ is the efficiency to identify
the primary lepton, $\varepsilon_{\mu^+ID}$ is the probability to
identify the muon produced at the decay vertex and to measure the
charge for muons reaching the magnet spectrometer,
$\varepsilon_{charge}$ is the probability to correctly identify the
charge of the daughter muon and $\varepsilon_{det}$ is the probability
to detect the charm decay. The latter efficiency includes the trigger
efficiency, a fiducial volume cut (emulsions cannot be scanned up to
the edge and there are small cracks between bricks, see
Ref.~\cite{proposal} for more details), the probability to correctly
identify with electronic detectors the brick where the interaction
occurred, the efficiency of the emulsion tracking algorithm in
reconstructing the interaction vertex in the ECC.

In our case only prompt unoscillated $\nu_e$ contribute to the
background, although this background is highly suppressed by the fact
that the charge of the muon produced at the charm decay vertex is
positive. Therefore, it contributes to the background only if
the charge of the muon is wrongly measured. By assuming an electron
identification efficiency of 97\% and the muon identification and
charge determination efficiencies coming from the algorithm described
in Section~\ref{muonsec}, we expect a background smaller than
$10^{-8}\times N_{l}^{CC}$.

\subsubsection{Anti-neutrino induced charm-production}
\label{sec:anticharm}
Charm-production from anti-neutrinos can be induced from both
unoscillated $\bar{\nu}_\mu$ and $\bar{\nu}_e$ from
$\bar{\nu}_\mu\rightarrow\bar{\nu}_e$ oscillations through the
reaction

$$\bar{\nu}_l N\rightarrow l^+ + C^- + X\, .$$

It is worth noticing that this background does not profit of the
charge measurement of the muon from charmed-hadron decay, being the charge of
the decay product negative.

The charm-production rate in anti-neutrino induced charm-production
has been estimated by using the approach discussed in Appendix~A. The
expected rate, normalized to CC interactions, are

$$\bar{R}_c(\nu_\mu) = (4.84\pm1.94)\%;\, \bar{R}_c(\nu_e) =
(4.36\pm1.48)\%\, .$$

Finally, we can write the expected number of events from this background as

$$\bar{N}_c = \bar{N}_{l}^{CC}\times \bar{R}_c(l)\times f_{C^+}\times
BR(C^-\rightarrow l^-)\times\overbrace{(1-\bar{\varepsilon}_{lID})
\times\bar{\varepsilon}_{charm}\times\varepsilon_{\mu^-ID}}
^{\varepsilon_\mu}\times\bar{\varepsilon}_{det}$$

where $\varepsilon_\mu$ gives the product of the probability that the
primary lepton is not identified, the decay is classified as long and
the secondary muon is reconstructed with the right charge. This
probability has been evaluated to be $1.7\times10^{-3}$ and it is
mildly dependent on the neutrino energy.

In our case only prompt unoscillated $\bar{\nu}_\mu$ contribute to the
background, which amounts to 

$$\bar{N}_c=3.7\times10^{-6}\times \bar{N}_{l}^{CC}\, .$$

\subsubsection{Events from $\tau^+\rightarrow\mu^+$ decays}
\label{tauplus}
A potential source of background comes from $\tau^+$, produced by
$\bar \nu_\tau$ from the leading $\bar{\nu}_\mu\rightarrow\bar{\nu}_\tau$
oscillations and undergoing to a charged-current interaction, decaying
into a positive muon with the charge wrongly measured.

From Tables~\ref{tab:lept732mm} and~\ref{tab:lept3000mm}, we see that
for $\theta_{13}=5^\circ$ and $\delta = 90^\circ$ the number of
$\tau^+$ is one order of magnitude larger than $\tau^-$. Therefore,
under the assumption that the $\tau$-decay detection efficiency is
independent of the lepton charge and that the probability to
misidentify the charge of the muon is of the order of $10^{-3}$, we
expect that for this particular value of ($\theta_{13},\delta$) the
background from this channel is a factor hundred smaller than the
observed number of signal events. Nevertheless, the dependence on the
oscillation parameters is properly taken into account.

\subsubsection{Muons wrongly matched to a hadron track}
Occasionally, a $\mu^+$ identified as $\mu^-$ in the electronic
detector is wrongly matched to a hadron track at the exit of the
vertex brick, as it is reconstructed by tracking with the emulsion ECC
brick. Because the kinematical analysis in the muonic channel is very
loose, without special care in the track matching such a mismatch
would result in a non negligible number of background events due to
hadron re-interactions.

In order to reduce this background, if there is an emulsion track with
an angular difference smaller than 50~mrad with respect to the one
matched to the ``electronic'' muon, the matching is flagged as
ambiguous and, in the presence of a kink, the event is not taken as a
$\tau$ candidate. This implies a few percent relative loss of
efficiency already accounted for in Section~\ref{expesigna}. Further
improvements in the background reduction come from the requirement to
have the charge of the muon correctly identified, the latter being not
present in the OPERA analysis discussed in Ref.~\cite{Guler:2001hz}.

The mismatch probability has been computed to be
$6\times10^{-4}$. On the other hand the probability that a hadron
undergoes to a re-interaction mimicking a kink topology which survives
the kinematical cuts has been computed with the FLUKA
package~\cite{fluka} to be $8\times10^{-4}$.
Therefore, the expected background from this source is

$$N_{mism}=7\times10^{-9}\times\bar{N}_{\mu}^{CC}\, .$$

\subsubsection{Decay in flight of $h^-$ and punch-through $h^-$}
\label{dif}

Neutral-current (NC) events with punch-through hadrons or particles
decaying in flight will dominate the scanning load of an OPERA-like
detector at the Neutrino Factory (see
Section~\ref{sec:scanning}). Moreover, if the hadron undergoes to an
elastic scattering in lead mimicking a kink topology, it will also
contribute to the silver channel background.  The probability of a
primary hadron to be reconstructed as a muon, reach the spectrometer
and have a reconstructed charge consistent with a silver candidate has
been computed by Monte Carlo simulation. It corresponds to an
efficiency of 0.6\%.  Folding the events with the probability of
undergoing large $p_T$ scattering in the first two lead sheets we
obtain an efficiency of $4.1 \times \ 10^{-6}$.  About 83\% of these
events are decay in flight. The other 17\% fraction is made up of
punch-through pions surviving the cut on the muon length.  Also
$\nu_e$~CC events contribute to this background if the primary
electron is unidentified. The background is

$$ N_{decay} = 4.1 \times 10^{-6} \times \varepsilon_{det} \times
\left[ \bar{N}^{NC}_{\mu}+ {N}^{NC}_{e} + (1-\varepsilon_e) \times
N^{CC}_{e} \right] \simeq 1.0 \times 10^{-6} \times
\bar{N}^{CC}_{\mu} $$

\noindent where $N^{NC}_{l}$ ($N^{CC}_{l}$) is the expected NC (CC)
rate for neutrinos of type $l$; $\varepsilon_e$ is the electron
efficiency and $\varepsilon_{det}$ is defined as in
Section~\ref{sec:nu_induced_charm}.

\subsubsection{Large-angle muon scattering}
\label{largemuon}
A muon can undergo to a large-angle scattering mimicking a muonic
decay of a short-lived particle. Potential sources of this background
are $\mu^-$ produced in CC interactions of $\nu_\mu$ from
$\nu_e\rightarrow\nu_\mu$ oscillations and $\mu^+$ from $\bar{\nu}_\mu$CC 
interactions, with a wrongly measured charge.

Extensive studies of this background have been carried out by the
OPERA Collaboration. A Monte Carlo simulation including the lead
form-factors gave a rate of muon scattering off 2~mm lead mimicking a
$\tau$ decay of
$0.2\times10^{-5}/N_\mu^{CC}$~\cite{proposal}. Recently, the analysis
of a dedicated measurement of large-angle scattering of 9~GeV muons in
lead plates became available. The corresponding measured rate is
$(0.6^{+0.7}_{-0.6})\times10^{-5}/N_\mu^{CC}$~\cite{Guler:2001hz}.  In
the OPERA proposal, a conservative assumption of $1 \times
10^{-5}/N_\mu^{CC}$ was made. Even in this case, the background from
this source is smaller than $\mathcal{O}(10^{-8})\times N_\mu^{CC}$

\subsubsection{Associated charm-production}
Another possible source of background is given by the associated
charm-production, in which two charmed hadrons are produced and one of
the two escapes the detection. The cross-section of this process has
been measured only by the E531 experiment and it turns out to be more than
one order of magnitude smaller than that for single
charm-production. In the following, we neglect this background.


\section{Estimate of the scanning load}
\label{sec:scanning}

We try here to give an estimate of the overall scanning load needed to
handle the silver and golden muon signals at the ECC detector and how
it scales with the detector mass.

There are several sources of events contributing to the scanning load.
Considering a $\mu^+$ in the storage ring, we can classify them in
four categories: signal events from $\nu_e \to \nu_\mu, \nu_\tau$
oscillations; background events from $\bar \nu_\mu \to \bar \nu_\mu,
\bar \nu_\tau \to \mu^+$, with misidentified muon charge (as discussed
in Sections \ref{tauplus} and \ref{largemuon}); background events from
both $\nu_e$ and $\bar \nu_\mu$ with charged charmed meson production
and unidentified primary lepton (as in Sections
\ref{sec:nu_induced_charm} and \ref{sec:anticharm}); background events
from $\nu_e$ and $\bar \nu_\mu$ NC interactions with punch-through
mesons misidentified as muons (Section \ref{dif}).  Whereas the signal
events depend on the mixing matrix parameters, the different sources
of background are mainly parameter-independent and can be estimated by
the knowledge of the $\nu_e$ and $\bar \nu_\mu$ fluxes at a given
distance from the source.

Consider a $\mu^+$ in the storage ring and the mixing matrix
parameters of Tab. \ref{tab:lept732mm}, namely $\theta_{13} = 5^\circ$
and $\delta = 90^\circ$, a baseline $L = 732$ km and the ``natural''
polarization ${\cal P}_{\mu^+} = 0$.  Any time a ``$\mu^-$'' (a
particle with the characteristics of a muon and with charge identified
as negative) is reconstructed by the electronic detector, we will have
to scan the emulsions of the brick where $\nu N$ interactions
occurred.  The first category of events are signal events: for a 1
Kton detector, we expect $\sim 310$ events from $\nu_e \to \nu_\mu$
oscillations and $\sim 30$ events from $\nu_e \to \nu_\tau$
oscillations with subsequent muonic tau decay, for the considered
mixing matrix parameters.  In the second category we find $\sim 2200$
events from $\bar \nu_\mu \to \bar \nu_\mu$ with misidentification of
the muon charge (however, after the emulsion scanning only large angle
muon scattering events will contribute to the background,
Section~\ref{largemuon}) and $\sim 2$ events from $\bar \nu_\mu \to
\bar \nu_\tau$ .  In the third category we find $\sim 80$ events from
$\bar \nu_\mu N \to \mu^+ C^-$ with non-observation of the primary
right-sign muon and correctly identified secondary wrong-sign muon
from the $C^-$ decay. All other kind of events in this category
($\nu_e \to l^- C^+$ and $\bar \nu_\mu \to e^+, \tau^+ C^-$) give
negligible contributions to the scanning load due to the oscillation
probability or to the $3 \times 10^{-3}$ factor for charge
misidentification.  These events will be scanned and something like 95
\% of them will be rejected when the primary lepton is identified
during the emulsion analysis.  Finally, events in the fourth category
give the most relevant contribution to the scanning load: these are NC
interactions of both $\nu_e$ and $\bar \nu_\mu$ where a pion or kaon
is identified as a muon (punch-through hadrons or their muonic decay).
Considering the neutrino fluxes in Tab.~\ref{tab:tab3} for the
polarization ${\cal P}_{\mu^+} = 0$ and $L = 732$ km, a factor 0.3
for the NC/CC cross-section ratio and the probability of $0.6 \times
10^{-2}$ for this process to occur (see Section~\ref{dif}) we have
about $\sim 4800$ events to be scanned.  Given the absence of a
physical kink these events will be rejected after emulsion analysis
except for hadron re-interactions.

In summary, for a 1 Kton mass detector located at $L = 732$ km from
the neutrino source, we estimate a total scanning load in 5 years
of data taking of less than $1 \times 10^4$. Notice that the
two dominant sources of scanning load, i.e. non-oscillated $\bar \nu_\mu$
CC interactions with misidentified muon charge and $\nu_e, \bar \nu_\mu$
NC interactions with punch-through mesons identified as muons, 
do not depend (to a great extent) on the mixing matrix parameters 
and therefore do not scale with $\theta_{13}$.
For a 5 Kton detector we expect less than $4 \times 10^4$ bricks to be scanned
in 5 years, a scanning load that seems technologically affordable. 

Finally, we remind that if the detector were to be located at $L =
3000$ km from the neutrino source, the scanning load will be reduced
by about a factor 15.

\section{Sensitivity to ($\theta_{13}, \delta$)}
\label{sec:refined}

The overall sensitivity achievable in the plane ($\theta_{13},
\delta$) is evaluated by combining golden muon events measured by a 40
Kton iron detector run with both beam polarities and silver muon
events observed with a 5 Kton OPERA-like detector. The golden muon
sample at the OPERA-like detector is not considered in the present
analysis since it has a worse signal/background ratio compared with
the iron detector. Eventually, only the $\mu^{+}$ polarity is taken
into account for the OPERA-like detector.  In the following, the
magnetized iron detector is located at a fixed baseline $L= 3000$ km,
while both $L=732$ and $L=3000$ km are considered as possible
baselines for the ECC detector.

\subsection{Signal and background uncertainties}
\label{sec:errors}

The total number of golden muons and background events expected for
the iron detector is reported in~\cite{Cervera:2000vy}, whereas
Section~\ref{sec:anasilver} describes our analysis of the silver muons
and backgrounds in the OPERA-like detector.  In the present section we
summarize the corresponding systematic uncertainties for both
detectors.  Although our estimates are based upon the present
experimental knowledge, whenever a definite experimental program
allowing an improvement is foreseen we will make use of the expected
results.

The main sources of systematic uncertainties for the "silver" signal
events in the OPERA-like detector are the knowledge of the emulsion
scanning efficiency and the cross-section ratio
$\sigma_\tau/\sigma_\mu$.  In the following we assume an overall
systematic uncertainty of 15\% which is consistent with the one used
in Ref.~\cite{proposal}.
 
For an OPERA-like detector three main background contributions (see
Section~\ref{sec:anasilver}) are present: the muonic decay of $\tau^+$
events from $\bar{\nu}_\mu\rightarrow\bar{\nu}_\tau$ oscillations, the
anti-neutrino induced charm-production and the decay in flight of
$h^-$ and punch-through $h^-$. The present knowledge on anti-neutrino
induced charm-production, as discussed in Appendix~A, is limited to
about 40\% because of lack of data. However, at the Neutrino Factory a
specific short-baseline program is foreseen~\cite{Mangano:2001mj},
collecting a sample of $\mathcal{O}(10^6)$ events induced by neutrinos
and anti-neutrinos with a charmed hadron in the final state. Such a
statistics will improve considerably the knowledge of the related
cross-sections. In the following we assume a conservative 10\%
systematic uncertainty on the charm background.  The uncertainty on
the background induced by $\tau^+$ decays is the same as the one on
the "silver" signal. The last background contribution is dominated by
the poor knowledge of the hadronization and of the hadronic
re-interaction processes. We conservatively assume a 50\%
uncertainty on this background source.

A preliminary study of the performance of a magnetized iron detector
for the $\nu_e\rightarrow\nu_\mu$ search at a Neutrino Factory was
presented in~\cite{Cervera:2000vy}. The main background sources were
pion and kaon decays and muonic decay of charmed hadrons.  A 20\%
resolution was assumed on the reconstructed neutrino energy, while no
systematic uncertainty was assigned to backgrounds and signal events
in the calculations. In order to obtain realistic estimates of the
experimental performance we try to include such uncertainties in the
present work. The systematic uncertainty on the expected number of
signal events is mainly related to the detection efficiency and to the
calibration, since the flux and the cross-section are known to a high
accuracy. In the following we assume a 10\% uncertainty on the signal,
taking into account the coarse granularity of the detector and the
absence of an {\em in situ} calibration beam line (used for instance
to calibrate and monitor the NuTeV calorimeter~\cite{Harris:1999uv}).
The dependence of the sensitivity on this uncertainty will be
discussed in the following sections. As far as the background is
concerned, we assume a 50\% systematic uncertainty, consistently with
the treatment of the OPERA-like detector, although the overall
sensitivity only mildly depends on this parameter.

\begin{figure}[htbp]
\begin{center}
\begin{tabular}{c}
\hspace{-0.3cm}
\epsfxsize12.5cm\epsffile{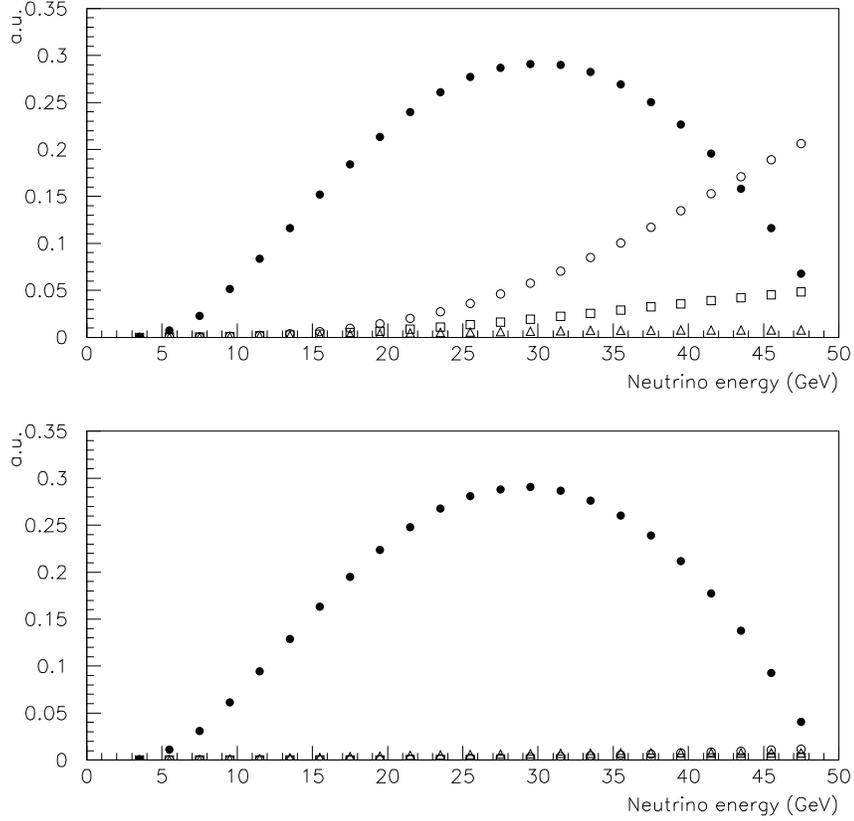}
\end{tabular}
\caption{\it Signal (black circles) and backgrounds for a 1 Kton
detector as a function of the neutrino energy at two different baselines: 
$L=730$ km (top panel), $L=3000$ km (bottom panel). 
Background from anti-neutrino induced charm-production (white circles), 
$\tau^+\rightarrow\mu^+$ decays (white triangles) and decay in flight of 
hadrons in $\bar{\nu}_\mu$ and ${\nu}_e$ (white squares) neutral-currents 
interactions are also shown. The signal
corresponds to $\theta_{13} = 5^\circ$ and $\delta = 90^\circ$.}
\label{fig:backvseve}
\end{center}
\end{figure}

\subsection{Statistical treatment}
\label{sec:stat}

Since signal and backgrounds have a different energy dependence (see
Fig.~\ref{fig:backvseve}), the number of expected events in both
detectors is divided into several energy bins. For the iron detector
we use five bins of constant width of 10 GeV~\cite{Cervera:2000vy,
Cervera:2000kp,Burguet-Castell:2001ez}.  An energy resolution of
$\Delta E / E = 20\%$ is used to smear the expected number of events
in each bin, resulting in a slightly lower signal/background
ratio. The number of signal and background events for 5 years data
taking and for both polarities in all bins is shown in
Tab.~\ref{tab:ironbins}.

\begin{table}[htb]
\centering
\begin{tabular}{|l|r|r|r|r|r|} \hline
 \multicolumn{1}{|c|}{\em Energy} & \multicolumn{1}{|c|}{\em Expected}
 & \multicolumn{4}{|c|}{\em Expected signal ($\theta_{13},\delta$)} \\
 \multicolumn{1}{|c|}{\em \& beam} & \multicolumn{1}{|c|}{\em back.} &
 ($1^\circ,0^\circ$) &
 ($1^\circ,90^\circ$) &
 ($5^\circ,0^\circ$) &
 ($5^\circ,90^\circ$) \\ \hline \hline 0$\div$10,
 $~\mu^+$ & 1.31 & 0.25 & 0.45 & 3.36 & 4.24 \\ 11$\div$20, $\mu^+$ &
 8.50 & 61.99 & 61.27 & 685.76 & 692.67 \\ 21$\div$30, $\mu^+$ & 2.43
 & 109.53 & 80.50 & 1122.08 & 984.82 \\ 31$\div$40, $\mu^+$ & 0.54 &
 81.18 & 56.60 & 866.17 & 726.54 \\ 41$\div$50, $\mu^+$ & 0.21 & 42.61
 & 26.97 & 422.18 & 366.95 \\ \hline\hline 0$\div$10, $~\mu^-$ & 2.90
 & 0.19 & 0.08 & 0.96 & 0.36 \\ 11$\div$20, $\mu^-$ & 8.79 & 29.15 &
 6.84 & 237.04 & 129.12 \\ 21$\div$30, $\mu^-$ & 7.51 & 59.08 & 17.38
 & 510.71 & 308.29 \\ 31$\div$40, $\mu^-$ & 4.37 & 44.18 & 15.26 &
 414.65 & 261.41 \\ 41$\div$50, $\mu^-$ & 1.44 & 24.20 & 8.43 & 215.90
 & 147.24 \\ \hline
\end{tabular} 
\caption{\em Number of background and golden muon signal events
expected in the iron detector at $L=3000$ km including our estimate of
the systematic effects. The binning definition refers to the
reconstructed energy .}
\label{tab:ironbins}
\end{table}

\begin{figure}[htb]
\begin{center}
\epsfig{file=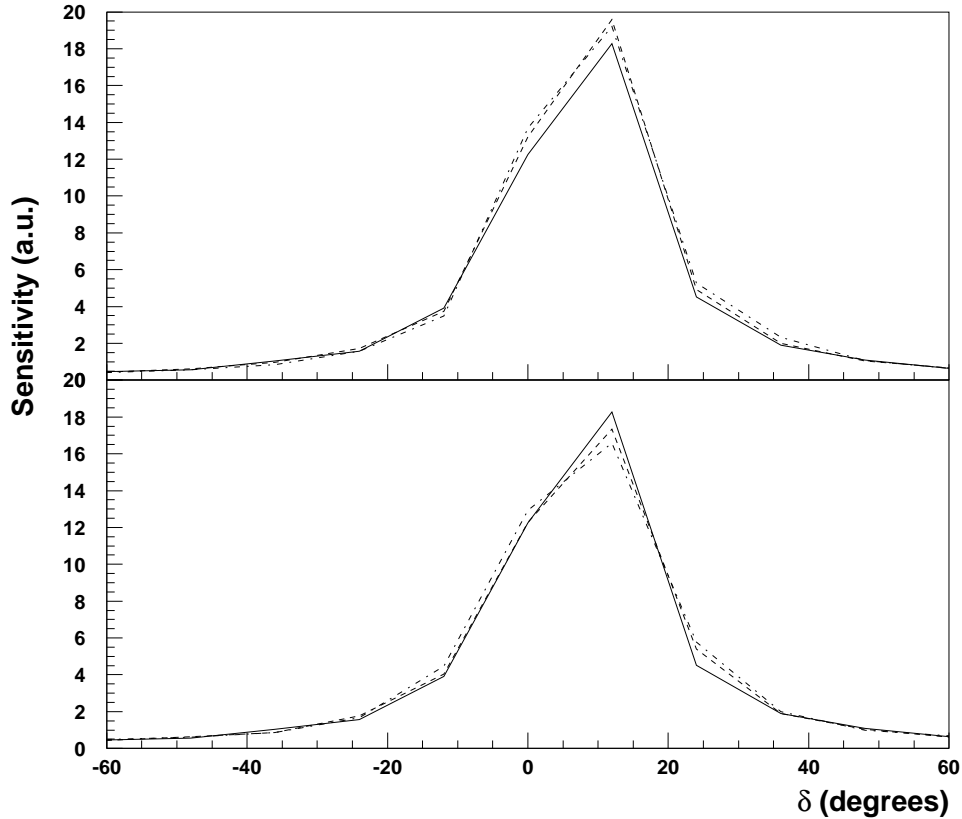,width=14cm}
\end{center}
\vspace{-0.5cm}
\caption{\em Sensitivity (arbitrary units) of the OPERA-like detector as a 
function of $\delta$ for $\theta_{13}<2^\circ$ and different bin 
configurations. The solid curve is the final choice with four bins. 
The dashed lines are obtained by changing the bin boundaries (top panel) 
and by increasing the total number of bins (bottom panel). Signal and 
background uncertainties are taken into account in the calculation.}
\label{fig:binsens}
\end{figure}

The choice of the binning for the OPERA-like detector is performed
according to the overall sensitivity to ($\theta_{13}, \delta$).  In
particular, since the number of expected events at small $\theta_{13}$
varies quadratically with $\theta_{13}$, the binning is optimized for
$\theta_{13}<2^\circ$ which is the critical region where few signal
events are expected and assuming for OPERA $L =732$~km. Several
configurations for both the total number of bins and their boundaries
are considered. For each binning definition, we evaluate the
sensitivity as the average 90\% CL upper limit obtained by a large
ensemble of experiments in the absence of a signal. The calculation is
performed for different values of $\delta$ in the range $\mid \delta
\mid <60^\circ$. The final configuration includes four bins and is the
one providing the best sensitivity as a function of
$\delta$. Fig.~\ref{fig:binsens}-a shows the sensitivity obtained by
changing the bin boundaries and Fig.~\ref{fig:binsens}-b the results
obtained with a different number of bins.  Tab.~\ref{tab:operabins}
summarizes the expected number of signal and background events in each
of the reconstructed energy bins.  The energy smearing has a larger
impact for the OPERA-like detector due to the strong energy-dependence
of the background (dominated by charm-production), thus increasing the
expected background at low energy. From Tab.~\ref{tab:operabins} it
can be seen that the OPERA-like detector has no sensitivity when both
$\theta_{13}\leq 1^0$ and $\mid \delta \mid < 15^\circ$. This region
partially corresponds to the central part of the curves in
Fig.~\ref{fig:binsens}.

\begin{table}[htb]
\centering
\begin{tabular}{|l|r|r|r|r|r|} \hline
 \multicolumn{1}{|c|}{\em Energy} & \multicolumn{1}{|c|}{\em Expected}
 & \multicolumn{4}{|c|}{\em Expected signal ($\theta_{13},\delta$)} \\
 \multicolumn{1}{|c|}{\em \& distance} & \multicolumn{1}{|c|}{\em
 back} & ($1^\circ,0^\circ$) &
($1^\circ,90^\circ$) &
($5^\circ,0^\circ$) &
($5^\circ,90^\circ$) \\ \hline \hline 0$\div$18,
 ~732 & 0.75 & 0.01 & 0.39 & 4.69 & 6.64 \\ 19$\div$29, 732 & 3.91 &
 0.02 & 0.89 & 9.73 & 13.81 \\ 30$\div$42, 732 & 9.37 & 0.01 & 0.93 &
 9.37 & 14.10 \\ 43$\div$50, 732 & 9.82 & $<$0.01 & 0.40 & 4.08 & 5.99
 \\ \hline \hline 0$\div$18, ~3000 & 0.13 & 0.14 & 0.15 & 5.32 & 5.41
 \\ 19$\div$29, 3000 & 0.49 & 0.11 & 0.50 & 10.02 & 11.70 \\
 30$\div$42, 3000 & 0.92 & 0.07 & 0.59 & 9.13 & 11.92 \\ 43$\div$50,
 3000 & 0.86 & 0.02 & 0.27 & 3.85 & 5.04 \\ \hline
\end{tabular} 
\caption{\em Number of background and "silver" signal events expected
in the OPERA-like detector at 732 km and 3000 km. The binning
definition refers to the reconstructed energy.}
\label{tab:operabins}
\end{table}

In order to evaluate the confidence regions which can be realistically 
deduced from the experimental apparatus, we simulate several sets of 
data corresponding to different ($\bar{\theta}_{13}, \bar{\delta}$) points.  
For each theoretical point and each energy bin, we throw the numbers 
of (observed) signal and background events from Poisson distributions 
with mean values corresponding to the expected ones. 
Furthermore, the mean Poisson values are smeared assuming a 
gaussian width corresponding to the quoted uncertainties 
(see Section~\ref{sec:errors}).  
Notice that, in principle, the procedure should be repeted many times 
to compute the average confidence intervals from a large ensemble of 
identical experiments. However, given the prohibitive computational time, 
we only perform two trials for each simulated point.   

The individual measurements from each of the 14 energy bins (4 bins in
the OPERA-like detector for $\mu^+$ and 5 bins in the iron detector
for each beam polarity) are then combined using the frequentist
approach of Ref.~\cite{Feldman:1997qc} to set 68.27\% ($1\sigma$),
90\% and 99\% confidence intervals on the reconstructed ($\theta_{13},
\delta$) parameters. The different bins are treated as independent. A
maximum likelihood fit is performed to the simulated data to extract
the signal content, through a global scan in the plane
($\theta_{13},\delta$).  The internal signal grid used for the fit has
a resolution of $\Delta \theta_{13} = 0.1^\circ$ and $\Delta \delta =
1^\circ$.

\subsection{Combining an OPERA-like detector at 732 km and an iron detector at 3000 km}
\label{sec:o732i3000} 

The numbers of expected events summarized in Tab.~\ref{tab:ironbins}
and Tab.~\ref{tab:operabins} show that the sensitivity of the
OPERA-like detector degrades considerably at small values of both
$\theta_{13}$ and $\delta$. However, its contribution to the overall
measurement is very relevant due to the complementary oscillation
pattern described in Section~\ref{sec:channel}. Therefore, the effect
of combining both "golden" and "silver" channels can be already seen
for $\theta_{13}>1^\circ$.
 
\begin{table}[htb]
\centering
\begin{tabular}{|c|r|r|} \hline
 \multicolumn{1}{|c|}{\em Confidence level} & \multicolumn{1}{|c|}{\em
   $\delta_{min}$} & \multicolumn{1}{|c|}{\em $\delta_{max}$} \\ & &
   \\ \hline \hline 68.27~\% & $60.8^0$ & $98.6^0$ \\ 90.00~\% &
   $49.9^0$ & $106.6^0$ \\ 99.00~\% & $29.8^0$ & $118.1^0$ \\ \hline
\end{tabular} 
\caption{\em Allowed regions on the $\delta$ parameter extracted from
the analysis of the simulated data for
$\bar{\theta}_{13}=1^\circ,\bar{\delta}=90^\circ$.  Both a 5 Kton
OPERA-like detector at $L =732$ km and a 40 Kton iron detector at $L =
3000$ km are considered. The best fit corresponds to
$\theta_{13}=0.9^\circ, \delta=80^\circ$.}
\label{tab:o732i3000cls}
\end{table}

\begin{figure}[htb]
\begin{center}
\epsfig{file=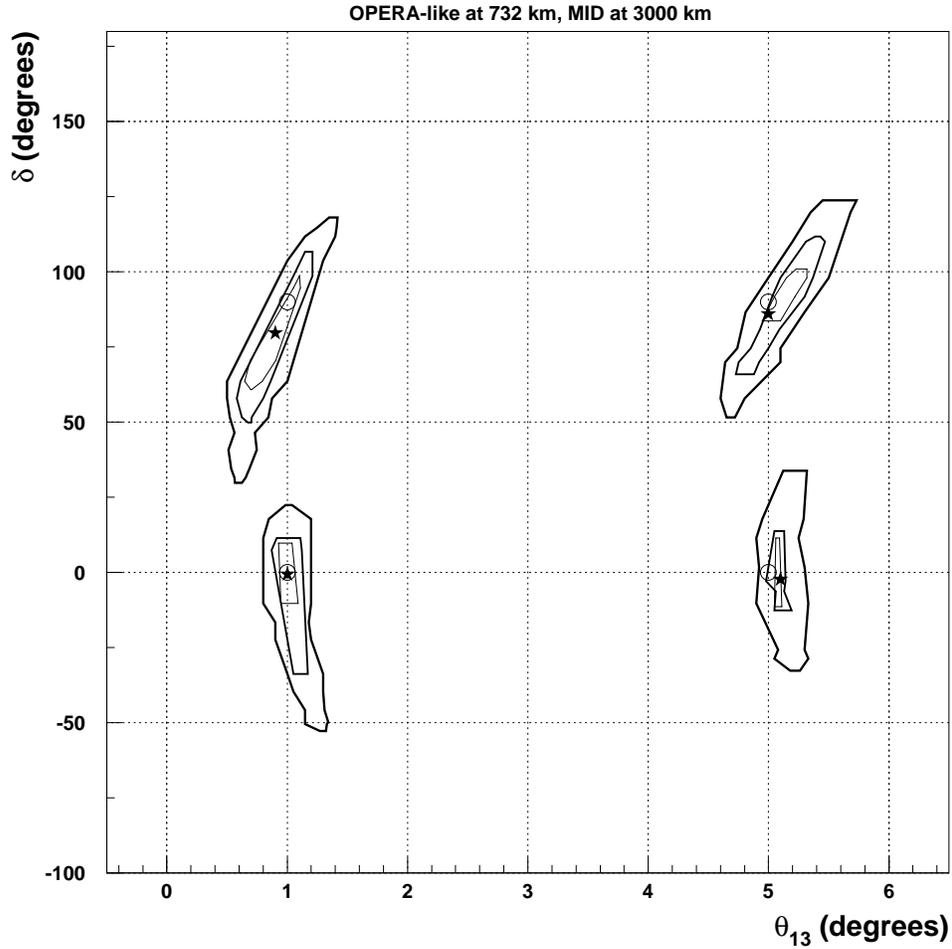,width=14cm}
\end{center}
\vspace{-0.5cm}
\caption{\em Expected 68.27\%, 90\% and 99\% confidence regions in the
($\theta_{13}, \delta$) plane corresponding to four simulated points.
Both an OPERA-like detector at 732 km and an iron detector at 3000 km
are combined. The stars denote the best fit points, while the open
circles are the true simulated points.}
\label{fig:o732i3000cls}
\end{figure}

Fig.~\ref{fig:o732i3000cls} shows the extracted 68.27\%, 90\% and 
99\% confidence regions in the ($\theta_{13}, \delta$) plane  
corresponding to four simulated points, if the OPERA-like detector 
is placed at $L = 732$ km from the neutrino source. 
Tab.~\ref{tab:o732i3000cls} summarizes the projection of the
confidence level contours onto the $\delta$ axis
for $\bar{\theta}_{13}=1^\circ, \bar{\delta}=90^\circ$. 
It can be noticed that all the displayed curves are connected.

Since the expected background is small ($<10$ events) in all bins, the
systematic uncertainties on this component have a minor effect on the
overall sensitivity. The same consideration applies to the "silver"
signal in the OPERA-like detector. However, systematic uncertainties
greater than 5\% on the golden signal can reduce the sensitivity of
the iron detector. This is particularly important at small values of
$\delta$. In our calculation we assumed a conservative 10\%
uncertainty, which could be further reduced by an accurate work on the
detector. Tab.~\ref{tab:sysiron} shows the effect of this parameter on
the 99\% confidence regions resulting from the simulated point
$\bar{\theta}_{13}=5^\circ, \bar{\delta}=0^\circ$.
 
\begin{table}[htb]
\centering
\begin{tabular}{|r|r|r|} \hline
 \multicolumn{1}{|c|}{\em Signal uncertainty}  & \multicolumn{1}{|c|}{\em $\delta_{min}$} & \multicolumn{1}{|c|}{\em $\delta_{max}$} \\ 
  \multicolumn{1}{|c|}{\em for iron detector} &  & \\ \hline \hline 
  0~\% & $-16.6^0$ & $20.5^0$ \\  
  5~\% & $-18.3^0$ & $23.5^0$ \\  
  10~\% & $-32.7^0$ & $33.8^0$  \\  
  20~\% & $-49.9^0$ & $61.7^0$ \\ \hline 
\end{tabular} 
\caption{\em Effect of the systematic uncertainty on the signal 
in the iron detector for the 99\% CL $\delta$ intervals obtained for 
$\bar{\theta}_{13}=5^\circ, \bar{\delta}=0^\circ$, when combining
the 5 Kton ECC detector at $L = 732$ km and the 40 Kton MID
at $L = 3000$. The best fit for the default configuration (i.e., with 
a 10\% systematic uncertainty on the golden muon signal at the MID) 
corresponds to $\theta_{13}=5.1^\circ, \delta=-2^\circ$.} 
\label{tab:sysiron}
\end{table}

From Tab.~\ref{tab:operabins} it can be seen that the sensitivity of
the OPERA-like detector at $L = 732$ km is limited by backgrounds,
mainly from anti-neutrino charm-production where the primary lepton is
not identified (Section~\ref{sec:anticharm}). The evaluation of this
component is based upon the present understanding of the OPERA
detector. As discussed in Section~\ref{muonsec}, a detailed $dE/dx$
analysis close to the end point of the tracks could further improve
the primary muon identificatin by about a factor two. The
corresponding efficiency loss is limited to few percents. When a
sizeable "silver" signal is expected, such a background reduction
would result in narrower confidence intervals, as can be seen from
Tab.~\ref{tab:o732i3000cls_0.5bkg}.

 
\begin{table}[htb]
\centering
\begin{tabular}{|c|r|r|r|r|} \hline
 \multicolumn{1}{|c|}{\em Confidence level}  & \multicolumn{2}{|c|}{\em Standard} & \multicolumn{2}{|c|}{\em Bkgnd$\times$0.5}  \\ 
  & \multicolumn{1}{|c|}{\em $\delta_{min}$} & \multicolumn{1}{|c|}{\em $\delta_{max}$} & \multicolumn{1}{|c|}{\em $\delta_{min}$} & \multicolumn{1}{|c|}{\em $\delta_{max}$} \\ \hline \hline 
  68.27~\% & $82.5^0$ & $101.5^0$ & $83.7^0$ & $102.6^0$ \\  
  90.00~\% & $65.4^0$  & $111.8^0$  & $81.4^0$  & $112.9^0$ \\  
  99.00~\% & $51.6^0$  & $123.8^0$  & $70.5^0$ & $127.8^0$ \\ \hline 
\end{tabular} 
\caption{\em Effect of a background reduction by a factor of two for the 
OPERA-like detector at 732 km and the simulated point 
$\bar{\theta}_{13}=5^\circ,\bar{\delta}=90^\circ$. The best fit for the 
standard configuration corresponds to $\theta_{13}=5.0^\circ, \delta=86^\circ$.} 
\label{tab:o732i3000cls_0.5bkg}
\end{table}

\subsection{Combining an OPERA-like detector and an iron detector both at 3000 km}
\label{sec:o3000i3000} 

\begin{figure}[htb]
\begin{center}
\epsfig{file=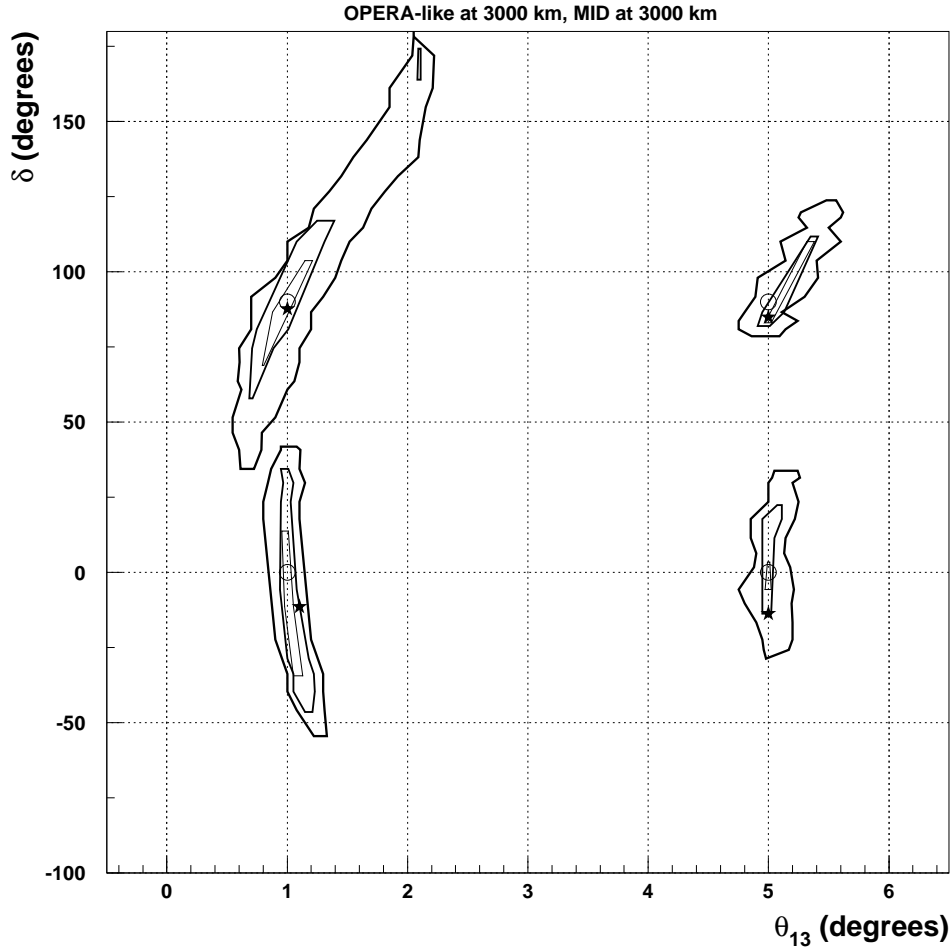,width=14cm}
\end{center}
\vspace{-0.5cm}
\caption{\em Expected 68.27\%, 90\% and 99\% confidence regions in the
($\theta_{13}, \delta$) plane corresponding to four simulated points.
Both an OPERA-like detector at 3000 km and an iron detector at 3000 km
are combined. The stars denote the best fit points, while the open
circles are the true simulated points.}
\label{fig:o3000i3000cls}
\end{figure}

A location of the OPERA-like detector at 3000 km would reduce considerably 
most background contributions (proportional to $1/L^{2}$), with the 
exclusion of the muonic decay of $\tau^+$ events from 
$\bar{\nu}_\mu\rightarrow\bar{\nu}_\tau$ oscillations. 
Tab.~\ref{tab:operabins} shows the corresponding increase in the 
signal/background ratio for the detection of the "silver" signal events.  

\begin{figure}[htb]
\begin{center}
\epsfig{file=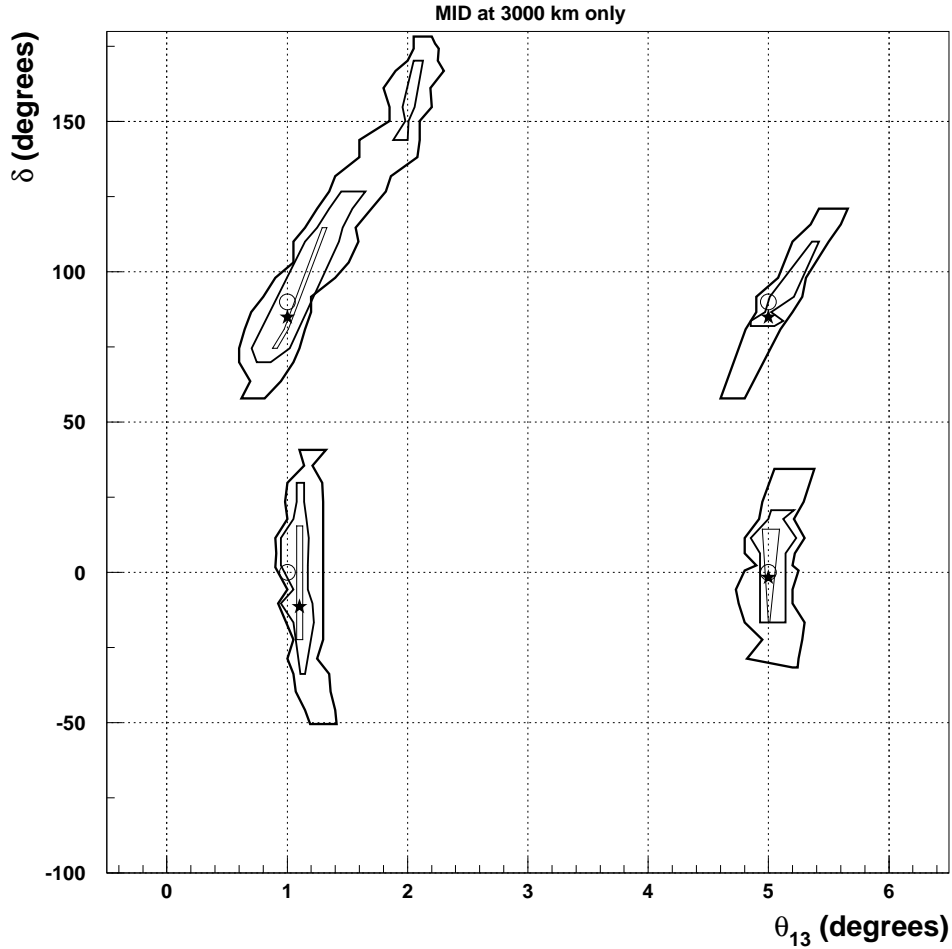,width=14cm}
\end{center}
\vspace{-0.5cm}
\caption{\em Expected 68.27\%, 90\% and 99\% confidence regions in the
($\theta_{13}, \delta$) plane corresponding to four simulated points.
Only an iron detector at 3000 km is considered, with two beam polarities. 
The stars denote the best fit points, while the open
circles are the true simulated points.}
\label{fig:i3000cls}
\end{figure}

In Fig.~\ref{fig:o3000i3000cls} we show the extracted 68.27\%, 90\%
and 99\% confidence regions in the ($\theta_{13}, \delta$) plane
corresponding to the same four simulated points of
Section~\ref{sec:o732i3000}. In this case, the curves extracted for
the point $\bar{\theta}_{13}=1^\circ,\bar{\delta}=90^\circ$ are not
fully connected, mainly due to the tiny signal contribution expected
from the "silver" channel (Tab.~\ref{tab:operabins}).  However, as
explained in the previous section, values of $\theta_{13}\sim 1^\circ$
are close to the intrinsic limit of the experimental sensitivity,
producing wide fluctuations on the number of observed
events. Actually, by increasing the number of simulated experiments
for $\bar{\theta}_{13}=1^\circ, \bar{\delta}=90^\circ$, in few cases
we observe the presence of disconnected curves also with the 732 km
baseline for the OPERA-like detector, regardeless of the additional
background reduction mentioned in Section~\ref{sec:o732i3000}.  For
larger values of $\theta_{13}$ the location of the OPERA-like detector
at 3000 km can produce some reduction in the size of the confidence
intervals. For comparison, in Fig.~\ref{fig:i3000cls} we give the 
corresponding results when only the magnetized iron detector 
at $L = 3000$ km is used (still considering two beam polarities). 

It is worth noting that the scanning load reduces as $1/L^2$ and
we do expect to scan few thousand events in this case. This
would allow some potential improvements:
\begin{itemize}
\item remove more than one brick per event. This would allow an
      increase of the signal detection efficiency by about 20\%
      (mainly due to an increase in the brick finding efficiency);

\item inclusion of the $\bar{\nu_{e}} \rightarrow \bar{\nu_{\tau}}$
      channel when the beam is run with opposite polarity.
      The background is larger by about a factor of two with
      respect to the $\nu_{e} \rightarrow \nu_{\tau}$ channel,
      while the signal is reduced by the same amount due to
      the cross-section of anti-neutrinos. Since the resulting
      background will be still small at 3000 km, this channel
      would slightly increase the sensitivity of the measurement in
      spite of the worse signal/background ratio.
\end{itemize}

Alternatively, one could also foresee an increase of the
mass of the detector and/or of the exposure.

\section{Conclusions}
\label{sec:concl}


It was previously shown~\cite{Cervera:2000kp} that looking for
``golden'' wrong-sign muons at the Neutrino Factory is the most
sensitive method to measure simultaneously $\theta_{13}$ and $\delta$.
One of the main problems concealed in this measurement was, however,
pointed out in~\cite{Burguet-Castell:2001ez}: due to correlations
between $\theta_{13}$ and $\delta$ in eq.~(\ref{eq:spagnoli}):
degenerate regions in the reconstructed ($\theta_{13}, \delta$)
parameter space occur in many cases, thus producing a scenario pretty
much similar to that of the solar neutrino puzzle before SNO and
KamLand. In particular, the appearance of different allowed regions in
the parameter space, often widely separated in the $\delta$ axis,
severely reduces the Neutrino Factory sensitivity to the CP-violating
phase.  In~\cite{Donini:2002rm} it was proposed the use of $\nu_e \to
\nu_\tau$ oscillation to solve the ($\theta_{13}, \delta$) ``intrinsic
ambiguity'' problem, taking advantage of the complementarity between
the $\nu_e \to \nu_\mu$ and $\nu_e \to \nu_\tau$ oscillation
probabilities (Section \ref{sec:channel}).

Whereas the ``golden'' channel can be studied using a coarsely grained
magnetized iron detector~\cite{Cervera:2000vy}, to profit of the
``silver'' channel we must use a detector of a different kind, capable
of separating the ``silver'' muon signal from the ``golden'' muon one,
by means of the different energy distribution of the final muons or by
looking for the $\tau$ decay vertex.  In~\cite{Donini:2002rm}, it was
decided to consider the second possibility looking for the ``silver''
muon signal with an OPERA-like detector, making use of the available
information on the ECC detector~\cite{proposal,Guler:2001hz}.
However, since the ``silver'' channel is strongly suppressed with
respect to the ``golden'' channel by the $\nu_\tau N$ cross-section
and by the $\tau \to \mu$ branching ratio (we are indeed dealing with
tens of events, at most), extreme care must be devoted to the
background treatment.  This is why we present in this paper a
dedicated analysis of backgrounds and efficiencies at an ECC
OPERA-like detector when dealing with the ``silver'' and ``golden''
channel at a Neutrino Factory. The outcome of this study, presented in
full detail in Section~\ref{sec:anasilver}, is that the three dominant
sources of background to the ``silver'' channel after emulsion
scanning, are (in order of importance): wrong-sign muons coming from
the decay of charged charmed mesons produced in combination with a
non-observed right-sign muon; punch-through mesons that mimick a
charged energetic lepton particle identified as a wrong-sign muon;
right-sign muons coming from the decay of a $\tau$ produced through
the leading oscillation $\nu_\mu \to \nu_\tau$, whose charge is
wrongly identified. Notice that the dominant sources of background
arise from non-oscillated $\nu_e$ and $\bar \nu_\mu$ that undergo CC
interactions (charmed mesons production) or NC interactions
(punch-through mesons): both of them, therefore, decreases like the
flux as $1/L^2$ and can thus be made negligible by locating the ECC
detector at a larger distance (as it was indeed the case for the
golden signal at the MID~\cite{Cervera:2000kp}).


In this paper we show that the intrinsic ambiguity problem is solved
for $\theta_{13} > 1^\circ$ when using a 40 Kton magnetized iron
detector, to deal with the ``golden'' muon signal, and a 5 Kton ECC
OPERA-like detector, to measure the ``silver'' muon signal. We also
include systematic effects in the treatment of the ``golden'' muon
sample at the MID, previously not considered. We present
(Section~\ref{sec:refined}) a refined statistical analysis of the
simulated data for a MID at $L = 3000$ km and an ECC either at $L =
732$ km or at $L = 3000$ km, taking advantage of the different energy
distribution of ``silver'' signal and backgrounds. Below $\theta_{13}
= 1^\circ$, the ``silver'' muon sample at the ECC detector becomes
statistically negligible. With both the MID and the ECC located at $L
= 3000$ km we expect a significant decrease in the background of the
``silver'' channel, as previously mentioned. This translates, for
$\theta_{13} > 1^\circ$, in a reduction of the confidence intervals
with respect to the configuration with the ECC at $L = 732$ km.


We have also carried out a rather detailed analysis of the foreseeable
scanning load at the 5 Kton ECC detector with the considered Neutrino
Factory beam ($1 \times 10^{20}$ useful $\mu^+$ decay in the storage
ring), showing that the total scanning load to deal with the
``silver'' and ``golden'' muon samples and the related backgrounds
($\sim 4 \times 10^4$ bricks to be scanned in five years) is
technologically affordable and does not represent a severe problem.

It must be stressed that in this paper we restricted ourselves to the
study of the ($\theta_{13},\delta$) intrinsic ambiguity, by fixing
$\theta_{23} = 45^\circ$ and by choosing a given sign for $\Delta
m^2_{atm}$ (in the hypothesis that more information on the three
neutrino spectrum will become available by the time the Neutrino
Factory will be operational).


We believe that solving the three ambiguities at the same time will
need the combination of different kinds of detectors and baselines: a
careful study of the required net of detectors is beyond the scope of
this paper and it will be presented elsewhere.



\newpage

\appendix
\section{Charm-production cross-section}
\label{charmcross}

\begin{figure}[htbp]
\begin{center}
\begin{tabular}{c}
\hspace{-0.3cm}
\epsfxsize12.5cm\epsffile{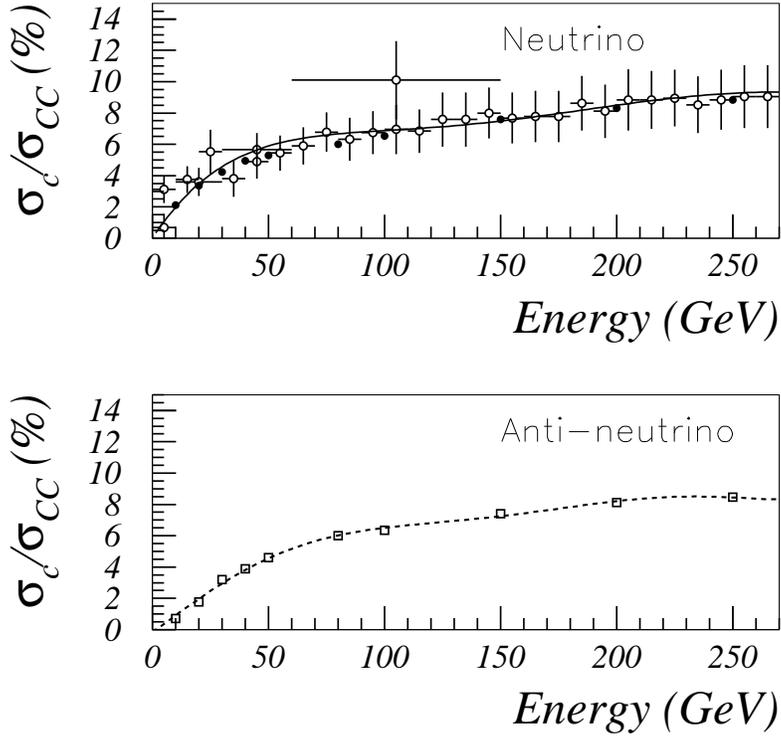}
\end{tabular}
\caption{\it Fraction of neutrino (anti-neutrino) total cross-section
that is associated with charm-production. Top: white circles show the
average inclusive charm-production cross-section as derived in
Ref.~\cite{DeLellis:2002mf} and the result of a fit to the data
(continous line); black circles show the predicted cross-section as
obtained by a leading-order calculation with
$m_c=1.31$~GeV~\cite{Conrad:1997ne}. Bottom: white squares show the
anti-neutrino induced charm-production as obtained by a leading-order
calculation with $m_c=1.31$~GeV~\cite{Conrad:1997ne}; the dashed line
show a fit to the point.}
\label{fig:charmcross}
\end{center}
\end{figure}

Neutrino and anti-neutrino induced single charm-production is
particularly interesting to study the strange-quark parton
distribution function and the threshold effect in the cross-section,
associated with the heavy quark production. Over the past 30 years,
many experiments have carried out these studies with complementary
techniques: calorimetry, bubble chambers and nuclear emulsions. These
data have been reviewed and combined statistically to extract a world
averaged single charm-production cross-section induced by neutrino in
Ref.~\cite{DeLellis:2002mf}. These results have been compared with
predictions of a leading-order calculation with $m_C=1.3$~GeV
(Fig.~\ref{fig:charmcross}) and the agreement is quite
good. Unfortunately, there are no inclusive measurements of the
anti-neutrino induced single charm-production. However, given the good
agreement between data and theoretical calculations obtained for
neutrinos, we use for our calculations the parametrization shown in
Fig.~\ref{fig:charmcross} which is based on the LO calculation at
$m_C=1.3$~GeV of~\cite{DeLellis:2002mf}.

\begin{thebibliography}{999}

\bibitem{Donini:2002rm}
A.~Donini, D.~Meloni and P.~Migliozzi,
Nucl.\ Phys.\ B {\bf 646} (2002) 321.

\bibitem{Pontecorvo:1957yb}
B.~Pontecorvo,
Sov.\ Phys.\ JETP {\bf 6} (1957) 429
[Zh.\ Eksp.\ Teor.\ Fiz.\  {\bf 33} (1957) 549];\\
Z.~Maki, M.~Nakagawa and S.~Sakata,
Prog.\ Theor.\ Phys.\ {\bf 28} (1962) 870;\\
B.~Pontecorvo,
Sov.\ Phys.\ JETP {\bf 26} (1968) 984;\\
V.~N.~Gribov and B.~Pontecorvo,
Phys.\ Lett.\ B {\bf 28} (1969) 493.

\bibitem{evidence_osc}
Y.~Fukuda {\it et al.}  [Super-Kamiokande Coll.],
Phys.\ Rev.\ Lett.\  {\bf 81}, 1562 (1998); \\
M.~Ambrosio {\it et al.}  [MACRO Coll.],
Phys.\ Lett.\ {\bf B517} (2001) 59; \\
B.T.~Cleveland {\it et al.},
Astrophys.\ J.\  {\bf 496}, 505 (1998); \\
J.N.~Abdurashitov {\it et al.}  [SAGE Coll.],
Phys.\ Rev.\ {\bf C60}, 055801 (1999); \\
W.~Hampel {\it et al.}  [GALLEX Coll.],
Phys.\ Lett.\ {\bf B447}, 127 (1999); \\
S.~Fukuda {\it et al.}  [Super-Kamiokande Coll.],
Phys.\ Rev.\ Lett.\  {\bf 86}, 5651 (2001); \\
Q.R.~Ahmad {\it et al.}  [SNO Coll.],
Phys.\ Rev.\ Lett.\  {\bf 87}, 071301 (2001); \\
M.H.~Ahn {\it et al.}  [K2K Coll.],
Phys.\ Rev.\ Lett.\  {\bf 90} (2003) 041801.\\

\bibitem{KAMLAND}
K.~Eguchi {\it et al.}  [KamLAND Coll.],
Phys.\ Rev.\ Lett.\  {\bf 90} (2003) 021802.

\bibitem{Athanassopoulos:1998pv}
C.~Athanassopoulos {\it et al.}  [LSND Collaboration],
Phys.\ Rev.\ Lett.\  {\bf 81} (1998) 1774;\\
A.~Aguilar {\it et al.}  [LSND Collaboration],
Phys.\ Rev.\ D {\bf 64} (2001) 112007.

\bibitem{zimmermann}
E.~Zimmerman,~talk~at~NOON~2003,~http://www-sk.icrr.u-tokyo.ac.jp/noon2003/.

\bibitem{yanagisawa}
C.~Yanagisawa,~talk~at~NOON2003,~http://www-sk.icrr.u-tokyo.ac.jp/noon2003/.

\bibitem{Ahmad:2002jz}
Q.~R.~Ahmad {\it et al.}  [SNO Collaboration],
Phys.\ Rev.\ Lett.\  {\bf 89} (2002) 011301;
Phys.\ Rev.\ Lett.\  {\bf 89} (2002) 011302

\bibitem{MSW} 
L. Wolfenstein, 
Phys. Rev. {\bf D 17} (1978) 2369;
Phys. Rev. {\bf D 20} (1979) 2634;\\
S.P. Mikheyev and A. Yu. Smirnov, 
Sov. J. Nucl. Phys. {\bf 42} (1986) 913.

\bibitem{Fogli:2002au}
G.~L.~Fogli, E.~Lisi, A.~Marrone, D.~Montanino, A.~Palazzo and A.~M.~Rotunno,
Phys.\ Rev.\ D {\bf 67} (2003) 073002;\\
A.~Bandyopadhyay, S.~Choubey, R.~Gandhi, S.~Goswami and D.~P.~Roy,
Phys.\ Lett.\ B {\bf 559} (2003) 121;\\
J.~N.~Bahcall, M.~C.~Gonzalez-Garcia and C.~Pena-Garay,
JHEP {\bf 0302} (2003) 009;\\
M.~Maltoni, T.~Schwetz and J.~W.~Valle,
arXiv:hep-ph/0212129.  

\bibitem{chooz}
M.~Apollonio {\it et al.} [CHOOZ Coll.],
Eur.\ Phys.\ J.\ C {\bf 27} (2003) 331.

\bibitem{ICARUS} F.~Arneodo {\it et al.}  [ICARUS Coll.],
ICARUS-TM/2001-08 LNGS-EXP 13/89 add.2/01.

\bibitem{proposal}
M. Guler et al., OPERA Collaboration, CERN/SPSC 2000-028, SPSC/P318, 
LNGS P25/2000. 

\bibitem{MINOS} E.~Ables {\it et al.} [MINOS Coll.],
FERMILAB-PROPOSAL-P-875 \\ 
The MINOS detectors Technical Design Report, NuMI-L-337, October 1998.

\bibitem{JHF}
Y.~Itow {\it et al.}, KEK-REPORT-2001-4, arXiv:hep-ex/0106019.

\bibitem{komatsu}
M.~Komatsu, P.~Migliozzi and F.~Terranova,
J.\ Phys.\ G {\bf 29} (2003) 443.

\bibitem{minosIN}
M.~Diwan et al., NuMI-NOTE-SIM-0714.

\bibitem{Migliozzi:2003pw}
P.~Migliozzi and F.~Terranova,
arXiv:hep-ph/0302274.

\bibitem{Huber:2002mx}
P.~Huber, M.~Lindner and W.~Winter,
Nucl.\ Phys.\ B {\bf 645} (2002) 3;
Nucl.\ Phys.\ B {\bf 654} (2003) 3;\\
P.~Huber and W.~Winter,
arXiv:hep-ph/0301257.

\bibitem{Geer:1998iz}
S.~Geer,
Phys.\ Rev.\ D {\bf 57} (1998) 6989
[Erratum-ibid.\ D {\bf 59} (1998) 039903].

\bibitem{DeRujula:1998hd}
A.~De Rujula, M.~B.~Gavela and P.~Hernandez,
Nucl.\ Phys.\ B {\bf 547} (1999) 21.

\bibitem{Blondel:2000gj}
A.~Blondel {\it et al.},
Nucl.\ Instrum.\ Meth.\ A {\bf 451} (2000) 102.

\bibitem{Apollonio:2002en}
M.~Apollonio {\it et al.},
arXiv:hep-ph/0210192.

\bibitem{Dick:1999ed}
K.~Dick, M.~Freund, M.~Lindner and A.~Romanino,
Nucl.\ Phys.\ B {\bf 562} (1999) 29.

\bibitem{Barger:2000fs}
V.~Barger, S.~Geer and K.~Whisnant,
Phys.\ Rev.\ D {\bf 61} (2000) 053004.

\bibitem{Bueno:2000wb}
A.~Bueno, M.~Campanelli and A.~Rubbia,
Nucl.\ Phys.\ B {\bf 573} (2000) 27.

\bibitem{Cervera:2000kp}
A.~Cervera {\it et al.},
Nucl.\ Phys.\ B {\bf 579} (2000) 17 [Erratum-ibid.\ B {\bf 593} (2001)
731]; Nucl.\ Instrum.\ Meth.\ A {\bf 472} (2000) 403.

\bibitem{Burguet-Castell:2001ez} 
J.~Burguet-Castell, M.~B.~Gavela,
J.~J.~Gomez-Cadenas, P.~Hernandez and O.~Mena,
Nucl.\ Phys.\ B {\bf 608} (2001) 301;\\
J.~Burguet-Castell and O.~Mena,
arXiv:hep-ph/0108109.

\bibitem{Minakata:2001qm}
H.~Minakata and H.~Nunokawa,
JHEP {\bf 0110} (2001) 001.

\bibitem{Barger:2001yr}
V.~Barger, D.~Marfatia and K.~Whisnant,
Phys.\ Rev.\ D {\bf 65}, 073023 (2002).

\bibitem{Minakata:2003ca}
H.~Minakata, H.~Nunokawa and S.~Parke,
arXiv:hep-ph/0301210.

\bibitem{Minakata:2002jv}
H.~Minakata, H.~Sugiyama, O.~Yasuda, K.~Inoue and F.~Suekane,
arXiv:hep-ph/0211111;\\
P.~Huber, M.~Lindner, T.~Schwetz and W.~Winter,
arXiv:hep-ph/0303232.


\bibitem{Freund:2001ui}
M.~Freund, P.~Huber and M.~Lindner,
Nucl.\ Phys.\ B {\bf 615} (2001) 331.

\bibitem{Rubbia:2001pk}
A.~Rubbia,
arXiv:hep-ph/0106088.

\bibitem{Bueno:2001jd}
A.~Bueno, M.~Campanelli, S.~Navas-Concha and A.~Rubbia,
Nucl.\ Phys.\ B {\bf 631} (2002) 239.

\bibitem{Kajita:2001sb}
T.~Kajita, H.~Minakata and H.~Nunokawa,
Phys.\ Lett.\ B {\bf 528} (2002) 245.

\bibitem{Burguet-Castell:2002qx}
J.~Burguet-Castell, M.~B.~Gavela, J.~J.~Gomez-Cadenas, P.~Hernandez and O.~Mena,
Nucl.\ Phys.\ B {\bf 646} (2002) 301.

\bibitem{Cervera:2000vy}
A.~Cervera, F.~Dydak and J.~Gomez Cadenas,
Nucl.\ Instrum.\ Meth.\ A {\bf 451} (2000) 123.

\bibitem{doniniNOON}
A. Donini, talk at NOON2003, http://www-sk.icrr.u-tokyo.ac.jp/noon2003/.

\bibitem{Guler:2001hz}
M.~Guler {\it et al.}  [OPERA Collaboration],
CERN-SPSC-2001-025.

\bibitem{Feldman:1997qc}
G.~J.~Feldman and R.~D.~Cousins,
Phys.\ Rev.\ D {\bf 57} (1998) 3873.

\bibitem{Freund:2001pn}
M.~Freund,
Phys.\ Rev.\ D {\bf 64} (2001) 053003.

\bibitem{Kimura:2002hb}
K.~Kimura, A.~Takamura and H.~Yokomakura,
Phys.\ Lett.\ B {\bf 537} (2002) 86; Phys.\ Rev.\ D {\bf 66} (2002) 073005.

\bibitem{Minakata:2002qe}
H.~Minakata, H.~Nunokawa and S.~Parke,
Phys.\ Lett.\ B {\bf 537} (2002) 249.

\bibitem{earthmodel} A. M. Dziewonski and D.L. Anderson, Phys. Earth
Planet. Int. {\bf 25} (1981) 297.

\bibitem{Lipari:2002at}
P.~Lipari,
Nucl.\ Phys.\ Proc.\ Suppl.\  {\bf 112} (2002) 274.

\bibitem{Zardetto}
P.~Lipari, M.~Lusignoli, D.~Meloni and D.~Zardetto, in preparation.

\bibitem{Conrad:1997ne}
J.~M.~Conrad, M.~H.~Shaevitz and T.~Bolton,
Rev.\ Mod.\ Phys.\  {\bf 70} (1998) 1341.

\bibitem{Paolo} 
P.~Lipari, private communication.

\bibitem{Broncano:2002hs}
A.~Broncano and O.~Mena,
arXiv:hep-ph/0203052.

\bibitem{mcsnim} M. De Serio et al., {\it Momentum measurement by the
  angular method in the Emulsion Cloud Chamber}, Accepted for
  publication on Nuclear Instruments and Methods.

\bibitem{powell} C.F. Powell et al., {\it The study of elementary
  particles by the photographic method}, Pergamon Press (1959).

\bibitem{DeLellis:2002mf}
G.~De Lellis, A.~Marotta and P.~Migliozzi,
J.\ Phys.\ G {\bf 28} (2002) 713
[Erratum-ibid.\  {\bf G28} (2002) 1515].

\bibitem{fractions}
G.~De Lellis, F.~Di Capua and P.~Migliozzi,
Phys.\ Lett.\ B {\bf 550} (2002) 16.

\bibitem{Kayis-Topaksu:2002xq}
A.~Kayis-Topaksu {\it et al.}  [CHORUS Collaboration],
Phys.\ Lett.\ B {\bf 549} (2002) 48.

\bibitem{fluka}
A.~Fasso, A.~Ferrari, P.~R.~Sala and J.~Ranft,
SLAC-REPRINT-2000-117
{\it Prepared for International Conference on Advanced Monte Carlo for
  Radiation Physics, Particle Transport Simulation and Applications
  (MC 2000), Lisbon, Portugal, 23-26 Oct 2000}.

A.~Fasso, A.~Ferrari and P.~Sala,
SLAC-REPRINT-2000-116
{\it Prepared for International Conference on Advanced Monte Carlo for
  Radiation Physics, Particle Transport Simulation and Applications
  (MC 2000), Lisbon, Portugal, 23-26 Oct 2000}.

\bibitem{Mangano:2001mj}
M.~L.~Mangano {\it et al.},
arXiv:hep-ph/0105155.

\bibitem{Harris:1999uv}
D.~A.~Harris {\it et al.}  [NuTeV Collaboration],
Nucl.\ Instrum.\ Meth.\ A {\bf 447} (2000) 377.

\end{thebibliography}
\end{document}